\documentclass[
    aps,prb,twocolumn,
	groupedaddress,
	amsfonts,amssymb,amsmath,
	citeautoscript,longbibliography,
	letterpaper,nofootinbib
	]{revtex4-2}

\usepackage{graphicx,physics,hyperref,xcolor}

\DeclareMathOperator{\sgn}{sgn}
\begin{document}
\title{Chern Dartboard Superconductors}
\author{Rebecca Chan}
\author{Taylor L. Hughes}
\affiliation{Department of Physics and Institute for Condensed Matter Theory,
University of Illinois at Urbana--Champaign, Urbana,
IL 61801, USA}
\begin{abstract}
We investigate the interplay of particle-hole symmetry and sub-Brillouin zone (sBZ) topology by coupling a so-called Chern dartboard insulator (CDI) to a superconductor (SC) via the proximity effect. We dub the hybrid system, and equivalent intrinsically superconducting phases, a \emph{Chern dartboard superconductor} (CDSC). We show that a CDSC can have nontrivial sBZ topology if it arises from a CDI that has an even number of mirror symmetries $n$. On the other hand, particle-hole symmetry constrains a CDSC that arises from an odd-$n$ CDI to have trivial sBZ topology. However, we can circumvent this constraint for $n=1$ by inducing an FFLO-type pairing or shifting the CDI in momentum space, converting the mirror symmetry to a momentum-space nonsymmorphic mirror symmetry. With a superconducting pairing that preserves the (nonsymmorphic) mirror symmetries, even-$n$ CDIs and the shifted $n=1$ CDI can realize the minimal spinless phase that has a trivial total Chern number and nontrivial reduced Chern numbers. With a pairing that breaks the mirror symmetries, the hybrid system can realize phases that have nontrivial total and reduced Chern numbers, expanding the classification of phases that have sub-Brillouin zone (sBZ) topology. We also predict that some types of $n=2$ CDSCs inherit the quantized crystalline response of the $n=2$ CDI, providing experimentalists with a well-defined way to probe the CDSC. Our work motivates further exploration of sBZ topology, bulk topology, and quantized response.
\end{abstract}

\maketitle 
\section{Introduction}
Electronic topology plays a significant role in the classification and characterization of phases of matter. The topology of an electronic band is often characterized by topological invariants that depend on the Bloch wavefunctions over the entire Brillouin zone (BZ). A simple example is a 2D quantum Hall system characterized by the Chern number, i.e., the integral of the Berry curvature over the BZ~\cite{PhysRevLett.71.3697}. When the total Chern number vanishes, a finer topological classification can be achieved by computing Chern numbers of subsystems of the whole. For example, in the presence of a symmetry, e.g., spin-conservation or mirror symmetry, Chern numbers of each spin-component or each sector labeled by a mirror eigenvalue can reveal richer topological information than just the total Chern number~\cite{Sheng_QSH,Teo_mirror}. 

A recent investigation has introduced a new variant on the concept of a subsystem Chern number: the \emph{reduced} Chern number. This invariant is used to describe sub-Brillouin zone topology (sBZ)~\cite{cdi}, i.e., rather than being defined over the entire BZ, the reduced Chern number is the integral of the Berry curvature over a \emph{sector} of the full BZ. These sectors are non-overlapping regions of the BZ, typically related by a symmetry. In the initial work these sectors were defined by the high-symmetry lines of $n$ mirror symmetries~\cite{cdi}. Phases where the total Chern number is trivial but where the reduced Chern numbers are nontrivial were termed Chern dartboard insulators (CDIs). Depending on the number of mirror symmetries, the reduced Chern numbers can form a pattern resembling a domino ($n=1$, Fig.~\ref{fig:CDI}a), checkerboard ($n=2$, Fig.~\ref{fig:CDI}b), or dartboard ($n=3,4$) over the BZ. Interestingly, while the total Chern number of an insulator is proportional to the quantized Hall response~\cite{tknn}, the reduced Chern numbers were recently shown to generate quantized responses involving strain and electromagnetic effects~\cite{response}.

While the Chern number is most well-known in the context of topological insulators, nontrivial Chern numbers can also be used to classify 2D topological superconductors~\cite{readgreen,qhs_prox,ryuschn,Kitaev_2009}. This naturally leads us to ask if topological superconductors can support nontrivial reduced Chern numbers. While it seems plausible this should be the case, we will see that the mean-field, Bogoliubov-de-Gennes (BdG) description of topological superconductors satisfies a particle-hole transformation that constrains the sBZ topology. Drawing inspiration from earlier work that generates topological superconductivity by proximitizing a topological insulator~\cite{FuKane,qhs_prox}, a simple starting point to address our main question is to apply the superconducting proximity effect to an existing CDI. 
Explicitly, here we will couple a CDI to a superconductor and investigate the emergent phases and responses. As far as the topological phenomena are concerned, our results can be straightforwardly translated to systems that are intrinsically superconducting.  

Below we will split our discussion into cases having an even or odd number of mirror symmetries $n,$ as we will see that they behave differently in the presence of superconductivity. Intuitively, this distinction directly stems from the BdG particle-hole transformation flipping ${\bf{k}}\to -{\bf{k}}$ in the BZ. For example, consider  proximity-induced superconductivity in the $n=0$ base case, i.e., a Chern insulator, as in Ref.~\cite{qhs_prox}. In this example, the normal-state blocks that enter the BdG Hamiltonian, i.e., a Chern insulator and its particle-hole (PH) conjugate, have the \emph{same} total Chern number. Thus, even before adding superconductivity, the PH-doubled system has a nontrivial total Chern number, i.e., the superconducting Chern number (which counts chiral Majorana edge modes) is twice the Chern number of the initial Chern insulator (which counts chiral complex fermion edge modes). 

An important feature of this system is that the PH transformation takes the domain of the total Chern number, i.e., the full BZ, back to itself and does not generate any constraints on the Chern number. However, the PH transformation may constrain the reduced Chern numbers of a Chern dartboard superconductor (CDSC) in a nontrivial way because the sBZs may get mapped to each other under this transformation. To demonstrate this, let us consider the $n=1$ case where we have one mirror symmetry, say $M_y$ which takes $y\to -y.$ For this mirror symmetry the relevant sBZs are the upper and lower half planes of the BZ separated by the high-symmetry lines $k_y=0,\pi$ (we set the lattice constant to unity). Both the mirror symmetry and the PH transformation map one sBZ onto the other. Mirror symmetry requires the sBZ Chern numbers to have the same magnitude and opposite sign, while the standard PH transformation requires them to have the same magnitude and same sign. Hence, the reduced Chern numbers must vanish. Thus, the initial expectation is that even $n$ can support reduced Chern numbers while odd $n$ cannot. As we will see below this expectation can be circumvented if we choose different mirror symmetries or a different PH transformation.
 
Before moving on to the detailed discussion, we first summarize our results. We find that an even-$n$ CDI and its PH conjugate have identical reduced Chern numbers, so a proximitized superconducting system can have nontrivial sBZ topology similar to the $n=0$ case. For one choice of the proximity-induced pairing symmetry we find a state that has half the number of edge modes as the initial CDI. This CDSC is the minimal spinless state that has a trivial Chern number but nontrivial reduced Chern numbers, in analogy to the $p_x+ip_y$-like chiral superconductor for the $n=0$ case. For other pairing symmetries we obtain states that have both well-defined sBZ topology and a nontrivial \emph{total} Chern number, which expands the scope of sBZ topology. We also predict that some types of $n=2$ CDSCs exhibit quantized responses to crystalline deformations~\cite{response}. This response is an important finding because passing from a topological insulator to a topological superconductor usually complicates the quantization of electromagnetic topological responses connected to the Chern number. In our case the electromagnetic field is not involved and hence a well-defined, quantized response can persist.

In comparison, we find that odd-$n$ CDIs and their PH conjugates have opposite reduced Chern numbers, so the superconducting system has trivial reduced Chern numbers. However, we can obtain nontrivial sBZ topology for $n=1$ in two ways: (i) by modifying the $n=1$ CDI so that the sBZs are mapped to themselves under ${\bf{k}}\to-{\bf{k}}$ via the standard PH transformation, or (ii) by implementing a momentum-shifted PH transformation for a standard $n=1$ CDI by proximitizing with a Fulde-Ferrell-Larkin-Ovchinnikov (FFLO) state in which the Cooper pairs have a finite center-of-mass momentum. In proximity to an $s$-wave or extended $s$-wave SC, the CDSC obtained via either approach can exhibit sBZ topology analogous to the even-$n$ case. 

The remainder of our article is organized as follows. We first review the details of CDIs in Sec.~\ref{sec:CDIrev}. Next, in Sec.~\ref{sec:QAHrev}, we review the $n=0$ Chern insulator-SC system studied in Ref.~\cite{qhs_prox} to set up the method that we use to study the CDSCs. In Sec.~\ref{sec:n2} we consider the $n=2$ CDSC and explain how the pairing symmetry can generate phases that have nontrivial reduced and total Chern numbers and a well-defined, quantized response to strain. In Sec.~\ref{sec:n1} we examine the $n=1$ CDSC and how we circumvent the constraints of the conventional PH constraint to obtain nontrivial sBZ topology. Finally, we conclude in Sec.~\ref{sec:conc} with a brief summary of our results and future directions.

\section{Review of Chern dartboard insulator}
\label{sec:CDIrev}
\begin{figure}
    \centering
    \includegraphics[width=1.0\linewidth]{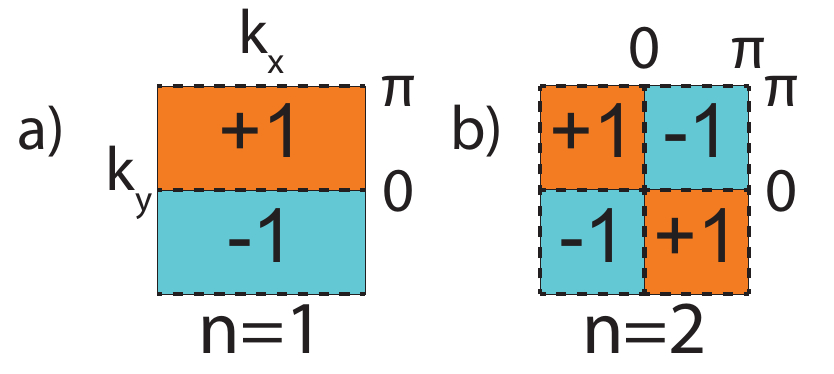}
    \caption{Sub-Brillouin zones (sBZs) and reduced Chern numbers for an $n=1$ and $n=2$ CDI. Dashed lines indicate the mirror high-symmetry lines (HSLs) demarcating the sBZs. An $n=1$ CDI has sBZs which are not fully bounded by HSLs.}
    \label{fig:CDI}
\end{figure}
Before discussing the hybrid CDSC system, we will review CDIs and sBZ topology. As introduced in Ref.~\cite{cdi}, Chern dartboard insulators belong to a class of 2D delicate topological systems that have trivial total Chern numbers $N$ and nontrivial quantized \emph{reduced} Chern numbers $N_r$. An $n$th-order CDI has $n$ mirror symmetries, \begin{equation}
    M_ih({\bf k})M_i^{-1}=h(R_i{\bf k}),\,\, i=1,\ldots n,
\end{equation}
which specify high-symmetry lines that divide the BZ into $2n$ sBZs. A reduced Chern number can be defined for each sBZ:
\begin{equation}
    N_r=\frac{1}{2\pi}\int_{\rm sBZ}d^2k\,\mathcal{F}({\bf k}),
\end{equation}
where $\mathcal{F}({\bf k})=\partial_{k_x}\mathcal{A}_y({\bf k})-\partial_{k_y}\mathcal{A}_x({\bf k})$ is the Berry curvature, and $\mathcal{A}_i({\bf k})=i\braket{u_{\bf k}}{\partial_{k_i}u_{\bf k}}$ is the Berry connection. It is not obvious that these integrals should be quantized, but mirror symmetry guarantees this. The mirror symmetries require the Berry curvature to vanish along the border of each sBZ. If we consider the $n=2$ case for example (see Fig.~\ref{fig:CDI}b), the boundary of each sBZ can be one-point compactified to a sphere $S^2$ since the Berry curvature is constant on the entire boundary. Since $S^2$ is a closed, compact manifold, the integral of the Berry curvature over each sBZ is quantized. Furthermore, in the presence of mirror symmetries we do not need to keep track of more than one of the $N_r$ since they are symmetry-related in a signed pattern as shown, for example, in Fig.~\ref{fig:CDI}.

In addition to the reduced Chern numbers, Ref.~\cite{response} showed that the $n=1$ and $n=2$ Chern dartboard insulators can also be characterized by their nonvanishing $n$th-order momentum-weighted Berry curvature multipole moments. These moments are well-defined only if all lower-order moments vanish---otherwise, shifting the origin of the BZ by a reciprocal lattice vector or gauge transformation can change the value of the multipole moment, rendering it unphysical. Using the framework of multipole moments, one can show that the $n=1$ CDI exhibits a mixed crystalline-electromagnetic response in which strain and dislocations bind charge density. Analogously, the $n=2$ CDI has a purely crystalline response in which translation dislocation defects bind momentum density. 

In what follows, we will couple CDIs to a superconductor via the proximity effect and examine the sBZ topology and response of the hybrid system. The method and analysis are inspired by the study of a Chern insulator in proximity to an $s$-wave SC~\cite{qhs_prox}. As mentioned, since the Chern insulator Hamiltonian has no mirror symmetries, it serves as the $n=0$ case where the single sBZ is the entire BZ. We will review the Chern insulator-SC hybrid system to lay the groundwork for discussing larger $n$. 

\section{Review of Chern Insulator-SC system (\MakeLowercase{n}=0)}\label{sec:QAHrev}
The minimal Chern insulator state has a single chiral edge state and a total Chern number $N=1$ where,
\begin{equation}
    N=\frac{1}{2\pi}\int_{\rm BZ}d^2k\,\mathcal{F}({\bf k}).
\end{equation} 
Ref.~\cite{qhs_prox} demonstrates that inducing an $s$-wave pairing in a Chern insulator state via the proximity effect immediately generates a topological superconducting (TSC) state that has a superconducting Chern number $\mathcal{N}=2N=2.$ The distinction between $N$ and $\mathcal{N}$ is that $|N|$ counts the number of protected complex chiral edge modes while $|\mathcal{N}|$ counts the number of protected chiral Majorana edge modes. It was further shown that a transition from this $\mathcal{N}=2$ state to an $\mathcal{N}=0$ trivial state can pass through a TSC phase characterized by $\mathcal{N}=1$ chiral Majorana edge modes. This $\mathcal{N}=1$ TSC phase has the fewest nonzero number of chiral Majorana edge modes and is the so-called \emph{minimal} fermionic topological state in 2D. Since the phenomenology for proximitizing CDIs follows closely that of the proximitized Chern insulator, we will now review the details carefully.  

We start with a simple Bloch Hamiltonian for a Chern insulator 
\begin{multline}
     h_{\rm CI}({\bf k})=\sin k_x\sigma_x+\sin k_y\sigma_y\\+(m+\cos k_x+\cos k_y)\sigma_z,\label{eq:QAH}
\end{multline}
where $\sigma_i$ are Pauli matrices representing spin. Here and throughout we set the lattice constant $a=1$. This Hamiltonian has $C_2$ symmetry represented by $C_2=\sigma_z$ but lacks mirror symmetries. We will focus on only $m>0$ because changing the sign of $m$ simply changes the sign of the Chern number, so the phase diagrams for $m>0$ and $m<0$ are the same under $N\rightarrow-N$. For $0<m<2$, the system is in an $N=1$ Chern insulating phase and has a single chiral mode on the edge. The bulk gap closes at ${\bf k}_0=(\pi,\pi)$ at the critical point $m=2$, and for $m>2$, the system is an $N=0$ trivial, normal insulator (NI) that has no protected gapless edge modes. 

The hybrid system of a Chern insulator and an $s$-wave superconductor can be described by the Bogoliubov-de Gennes (BdG) Hamiltonian \begin{equation}
    H_{\rm BdG}=\frac{1}{2}\sum_{\bf k}\Psi_{\bf k}^\dagger\begin{pmatrix}h_{\rm CI}({\bf k})&i\Delta\sigma_y\\-i\Delta\sigma_y&-h_{\rm CI}^*(-{\bf k})\end{pmatrix}\Psi_{\bf k},
\end{equation}
where $\Psi_{\bf k}=\begin{pmatrix}c_{{\bf k}\uparrow}&c_{{\bf k}\downarrow}&c_{-{\bf k}\uparrow}^\dagger&c_{-{\bf k}\downarrow}^\dagger\end{pmatrix}^T$ is the Nambu spinor. In line with Ref.~\cite{qhs_prox} we have chosen the pairing symmetry to be spin singlet $s$-wave. We will take $\Delta$ to be real and non-negative since the complex phase of $\Delta$ can be absorbed by a gauge transformation.  

The TSC phases of this model can be described by the superconducting Chern number $\mathcal{N}$. In the limit $\Delta=0$ the BdG Hamiltonian is the direct sum of the Chern insulator Hamiltonian in Eq. \ref{eq:QAH} and its PH conjugate. The Chern numbers $N_p$ and $N_h$ for the particle and hole states are identical for this model because charge conjugation changes neither the Berry curvature nor the integration region in momentum space. The superconducting Chern number $\mathcal{N}=N_p+N_h$ is then twice the Chern number of the Chern insulator system in the $\Delta=0$ limit: $\mathcal{N}=2$ when $0<m<2$ and $\mathcal{N}=0$ when $m>2$. In terms of edge states, the values of the Chern numbers imply that the complex chiral edge mode splits into two chiral Majorana edge modes as expected. 

We can now construct the phase diagram in the vicinity of the Chern-NI critical point as a function of $(m, \Delta)$ by block-diagonalizing the BdG Hamiltonian using \begin{equation}
    H_{\rm BdG}=M^{-1}\tilde H_{\rm BdG}M
\end{equation}
where \begin{equation}
    M=\frac{1}{2}\begin{pmatrix}
       -2\frac{d_z-\Delta}{d_x+id_y}&-\frac{d_x-id_y}{d_x+id_y}&\frac{d_x-id_y}{d_x+id_y}&2\frac{d_z-\Delta}{d_x+id_y}\\
       -1&0&0&1\\
       0&1&1&0\\
       1&0&0&1
    \end{pmatrix},
\end{equation}
and $d_{x,y,z}$ are the momentum-dependent coefficients of the Pauli matrices $\sigma_{x,y,z}$ in Eq. \ref{eq:QAH}. The block-diagonalized Hamiltonian is
\begin{equation}
    \tilde H_{\rm BdG}=\frac{1}{2}\sum_{\bf k}\tilde\Psi_{\bf k}^\dagger\begin{pmatrix}h_+&\\&-h_-^*(-{\bf k})\end{pmatrix}\tilde\Psi_{\bf k},\label{eq:blockd}
\end{equation}
where \begin{align}
    &h_\pm({\bf k})=h_{\rm CI}({\bf k})\pm\Delta\tilde\sigma_z\nonumber\\&=\sin k_x\tilde\sigma_x+\sin k_y\tilde\sigma_y
    +(m\pm\Delta+\cos k_x+\cos k_y)\tilde\sigma_z\nonumber\\
    &\equiv\sum_{i=x,y,z}d_i^\pm({\bf k})\tilde\sigma_i,
    \label{eq:hpmQAH}
\end{align} where $\tilde\sigma_i$ are Pauli matrices in the transformed basis. 

Hence, after the transformation Eq.~\ref{eq:hpmQAH} shows that the BdG Hamiltonian is equivalent to the direct sum of two Chern insulators that have mass parameters that are relatively shifted by the superconducting pairing: $m\rightarrow\tilde m=m\pm\Delta$. The Chern number of the superconducting system is the sum of the Chern numbers of $h_+$ and the PH conjugate of $h_-$. Since $h_-$ and its PH conjugate have identical Chern numbers, we have $\mathcal{N}=N_++N_-$, where $N_\pm$ are the Chern numbers of $h_\pm$.

We are now in a position to determine the phase boundaries near the Chern-NI critical point $(m,\Delta)=(2,0)$ for nonzero $\Delta$. Because topological invariants cannot change without closing the bulk gap, the phase boundaries can be determined by identifying the gapless points in the $(m,\Delta)$ plane. Both $h_\pm$ are gapless at ${\bf k}_0=(\pi,\pi)$ at the Chern-NI critical point, i.e., $d_{x,y,z}^\pm({\bf k}_0)=0$. Since $\Delta$ modifies only the mass parameter of $h_\pm$, the coefficients $d_{x,y}^\pm({\bf k}_0)$ still vanish, but $d_z^\pm({\bf k}_0)=m\pm\Delta-2$ vanishes only when $\Delta=\mp(m-2)$. Thus, the lines $\Delta=\mp(m-2)$ indicate where the Chern numbers of $h_\pm$ change, defining the phase boundaries that originate from the Chern-NI critical point at $\Delta=0$. 

Now that we know where the phase boundaries lie, we can characterize the gapped superconducting phases. When $\Delta<m-2$, both $h_+$ and $h_-$ have trivial Chern numbers and the system is in a trivial $\mathcal{N}=0$ phase. This superconducting phase is adiabatically connected to the trivial insulator phase in the $\Delta=0$ limit. When $\Delta<-m+2$, both $h_+$ and $h_-$ have nontrivial Chern numbers and the model is in an $\mathcal{N}=2$ TSC phase. If we smoothly reduce $\Delta$ to zero, the system becomes equivalent to two copies of the Chern insulator state, one of which is redundant because of the PH doubling. Between the $\mathcal{N}=2$ and $\mathcal{N}=0$ phases, when $\Delta>|m-2|$, $N_+=0$ while $N_-=1$, so the system is in an $\mathcal{N}=1$ TSC phase, and is not adiabatically connected to an insulating phase.

The phase diagram can also be understood through the evolution of the edge states. As mentioned above, the insulating Chern number $N$ counts the number of chiral fermionic edge modes, and the superconducting Chern number $\mathcal{N}$ counts the number of chiral Majorana edge modes. Each complex chiral mode that exists in the insulating phase can be trivially rewritten as two co-propagating chiral Majorana fermions. In the absence of superconductivity, the two Majorana modes are constrained to evolve together. As such, the two Majorana modes that make up the complex chiral edge mode of a Chern insulator must be degenerate and must be annihilated simultaneously at the Chern-NI critical point when $\Delta=0$. 

In contrast, turning on superconductivity via finite $\Delta$ can lift the degeneracy and the two chiral Majorana modes can evolve independently. As $\Delta$ increases, one of the Majorana modes becomes less localized (as controlled by the inverse gap size $(\sim(m-\Delta)^{-1})$ until it spreads fully into the bulk and annihilates with a counter-propagating partner on the opposite edge at the phase transition when $m=\Delta$. The resulting $\mathcal{N}=1$ state has one chiral Majorana mode on the edge, half the number of chiral Majorana edge modes as the Chern insulating state. This intermediate TSC phase is not equivalent to a fine-tuned limit of an insulating phase like the $\mathcal{N}=0,2$ phases. Since there is no topologically nontrivial 2D fermionic state that has fewer edge modes or a smaller superconducting Chern number, the $\mathcal{N}=1$ state is the minimal topological state in 2D.

In summary, since the Chern insulator state and its PH conjugate have identical Chern numbers, inducing superconductivity in the Chern insulator generates phases that have nontrivial superconducting Chern numbers. The transition from the $\mathcal{N}=2$ phase to the $\mathcal{N}=0$ phase passes through an intermediate $\mathcal{N}=1$ phase for finite $\Delta$. We note that an $s$-wave spin singlet pairing was sufficient to generate an $\mathcal{N}=1$ intermediate phase here. In order for superconductivity to open a gap at the Chern-NI critical point when $m=2$, $\Delta({\bf k})$ must be nonzero at ${\bf k}_0=(\pi,\pi)$. For the degrees of freedom in our model, only a spin singlet pairing can be nonzero at this point in the Brillouin zone due to Fermi statistics. 

We now want to use the procedure described in this section to investigate how the BdG PH transformation constrains the reduced Chern numbers of an $n$th-order CDI. We argued above that even and odd $n$ behave differently because the PH transformation may change the sBZ by taking ${\bf k}\rightarrow-{\bf k}$. Since the sBZs at ${\bf k}$ and $-{\bf k}$ have identical reduced Chern numbers for an even-$n$ CDI, an even-$n$ CDI and its PH conjugate should also have identical reduced Chern numbers. On the other hand, for an odd-$n$ CDI, the sBZs at ${\bf k}$ and $-{\bf k}$ have opposite reduced Chern numbers, so an odd-$n$ CDI and its PH conjugate should have opposite reduced Chern numbers. We will study even-$n$ CDIs first, as they should be higher-$n$ generalizations of this $n=0$ case. 

\section{Chern Dartboard Superconductor for \MakeLowercase{n}=2}\label{sec:n2}
\begin{figure}
    \centering
    \includegraphics[width=0.88\linewidth]{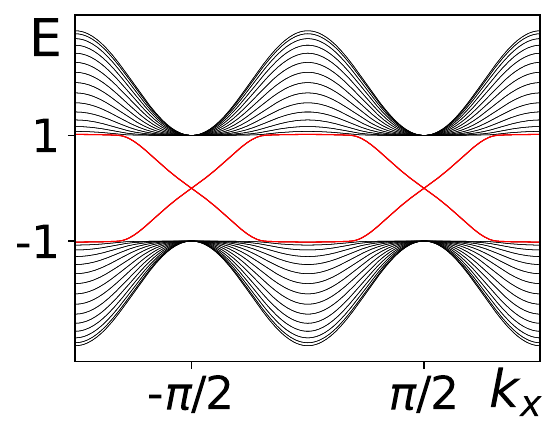}
    \caption{The $n=2$ CDI has doubly degenerate counter-propagating gapless edge modes. The energy spectra from 0 to $\pi$ and $-\pi$ to 0 are identical, as required by the mirror symmetries. }
    \label{fig:n2edge}
\end{figure}
In this section we focus on an $n=2$ CDI that has two orthogonal mirror symmetries $M_{x,y}$ that divide the BZ into four quadrants bounded by high-symmetry lines $k_{x,y}=0,\pi$, as shown in Fig.~\ref{fig:CDI}b. A reduced Chern number $N_r^{\pm\pm}$ can be defined for each of these quadrants, where $\pm\pm$ labels the sign of $k_x$ and $k_y$ within the quadrant. For example, $N_r^{+-}$, the reduced Chern number for the quadrant with $0<k_x<\pi$ and $-\pi<k_y<0$, is defined as
\begin{equation}
    N_r^{+-}=\frac{1}{2\pi}\int_0^\pi dk_x\int_{-\pi}^0dk_y\,\mathcal{F}({\bf k}).
\end{equation}  
Since a mirror operation changes the sign of the Berry curvature, the reduced Chern numbers for sBZs related by mirror symmetry have opposite signs: \begin{equation}
    N_r^{++}=N_r^{--}=-N_r^{+-}=-N_r^{-+}.
\end{equation} 
We can then define $N_r^a\equiv N_r^{++}=N_r^{--}$ and $N_r^b\equiv N_r^{+-}=N_r^{-+}$.

A prototypical Bloch Hamiltonian for an $n=2$ CDI is
\begin{multline}
    h_2({\bf k})=\sin k_x\sin2k_y\tau_x+\sin2k_x\sin k_y\tau_y\\+(m+\cos2k_x+\cos2k_y)\tau_z,\label{eq:n2CDI}
\end{multline}
where $\tau_i$ are Pauli matrices representing spinless $s$ and $d_{xy}$ orbitals, and $M_{x,y}=\tau_z$. For $0<m<2$, the model is characterized by $N_r^a=-N_r^b=-1$ and a Berry curvature quadrupole moment $Q_{xy}=2\pi^2$, where \begin{equation}
    Q_{xy}=\frac{1}{\pi}\int_{\rm BZ}d^2k\,k_xk_y\mathcal{F}({\bf k}).\label{eq:qxy}
\end{equation}
The trivial total Chern number and nontrivial reduced Chern numbers manifest through the presence of two degenerate counter-propagating modes on the edge, as shown in Fig.~\ref{fig:n2edge}. At $m=2$, the gap closes at four Dirac points located at the sBZ centers at ${\bf k}_{\pm\pm}=(\pm\pi/2,\pm/2)$. For $|m|>2$, the system is a trivial, normal insulator characterized by $N_r^a=N_r^b=0$, $Q_{xy}=0$, and no gapless edge modes. The bulk spectrum is gapless for $-2\le m\le0$. 

\subsection{General Pairing Symmetry}
When a generic pairing potential $\hat\Delta=\Delta({\bf k})\tau_i$ is induced in the $n=2$ CDI, the resulting $n=2$ Chern dartboard superconductor can be described by the BdG Hamiltonian \begin{equation}
    H_{\rm BdG}=\frac{1}{2}\sum_{\bf k}\Psi_{\bf k}^\dagger\begin{pmatrix}h_2({\bf k})&\Delta({\bf k})\tau_i\\\Delta^\dagger({\bf k})\tau_i&-h_2^*(-{\bf k})\end{pmatrix}\Psi_{\bf k},\label{eq:n2BdG}
\end{equation}
where $\Psi_{\bf k}=\begin{pmatrix}c_{{\bf k}s}&c_{{\bf k}d_{xy}}&c_{-{\bf k}s}^\dagger&c_{-{\bf k}d_{xy}}^\dagger\end{pmatrix}^T$ is the Nambu spinor. In the $\Delta=0$ limit, the BdG Hamiltonian is a direct sum of the CDI Hamiltonian in Eq. \ref{eq:n2CDI} and its PH conjugate. Since the reduced Chern numbers for the particle and hole states are equal, the superconducting reduced Chern number $\mathcal{N}_r=N_{p,r}+N_{h,r}$ is twice the reduced Chern number of the CDI model: $\mathcal{N}_r=\pm2$ when $0<m<2$ and $\mathcal{N}_r=0$ when $m>2$. 

To study the superconducting system for finite $\Delta$, we block-diagonalize the BdG Hamiltonian in Eq. \ref{eq:n2BdG} using the pairing-dependent basis transformations in Appendix \ref{sec:n2basis}. The block-diagonalized Hamiltonian takes the form of Eq. \ref{eq:blockd} where 
\begin{equation}
    h_\pm({\bf k})=h_2({\bf k})+|\Delta({\bf k})|\cdot
    \begin{cases}
    +\tilde\tau_0,&i=y\\
    \pm\tilde\tau_x,&i=z\\
    \pm\tilde\tau_y,&i=0\\
    \pm\tilde\tau_z&i=x\\
    \end{cases}.
\end{equation}
$\tilde\tau_i$ represent the transformed basis. In this transformed basis, mirror and $C_2$ symmetries are represented by $M_x=M_y=\tilde\tau_z$ and $C_2=M_x M_y=\tilde\tau_0$. The BdG Hamiltonian is thus equivalent to the direct sum of two $n=2$ CDIs that have $\Delta$-dependent modifications. 

Inspired by the $n=0$ Chern insulator case, we can investigate whether the transition from the $\mathcal{N}_r=\pm2$ phase to the $\mathcal{N}_r=0$ phase can pass through an intermediate superconducting phase characterized by nontrivial superconducting reduced Chern numbers that have unit magnitude. Just as in the previous section, we expect that such an intermediate phase would arise by tuning away from the $m=2$ critical point for nonzero $\Delta.$ This would also imply that such a phase is not adiabatically connected to a CDI or normal insulator phase in the $\Delta=0$ limit. Intuitively, this phase should emerge when $h_\pm$ have different reduced Chern numbers. We therefore examine how the matrix structure of the pairing affects the Berry curvature of $h_\pm$. 

First, consider a $\tilde\tau_0$ pairing structure which represents an overall energy shift. An energy shift does not affect the Bloch functions or the Berry curvature, so for this case $h_\pm$ always have the same reduced Chern numbers and the superconducting system can realize only $\mathcal{N}_r=\pm2,0$ phases. In contrast, a $\tilde\tau_x$ or $\tilde\tau_y$ term shifts the gap closing points in the BZ and allows the Berry curvature of $h_\pm$ to vary along the borders of the sBZs, thus breaking the one-point compactification of the sBZ. Hence, for these terms $h_\pm$ cannot be characterized by well-defined reduced Chern numbers. Finally, a $\tilde\tau_z$ term modifies the mass parameter $m\rightarrow m\pm|\Delta({\bf k})|$. This modification resembles the $n=0$ case, and we show below that the shifted mass parameter can generate phases for which $h_\pm$ have different reduced Chern numbers. Based on this analysis we will focus on pairings that have a $\tilde\tau_z$ matrix structure (which was derived from the untransformed pairing matrix structure $\tau_x$). 

We have considered how the matrix piece affects the possible CDSC phases, so we will move on to examining how the momentum dependence of $\Delta({\bf k})$ affects the reduced Chern numbers. In particular, let us consider the phases in the vicinity of the CDI-NI critical point $(m,\Delta)=(2,0)$. Explicitly, \begin{align}
    h_\pm({\bf k})&=\sin k_x\sin2k_y\tilde\tau_x+\sin2k_x\sin k_y\tilde\tau_y\nonumber\\
    &\quad+(m\pm|\Delta({\bf k})|+\cos2k_x+\cos2k_y)\tilde\tau_z\nonumber\\
    &\equiv\sum_{i=x,y,z}d_i^\pm({\bf k})\tilde\tau_i.\label{eq:hpm2}
\end{align}
A reduced Chern number can change only if the gap closes in the respective sBZ. Hence, to determine the phase diagram via the reduced Chern numbers we must identify the parameters $(m,\Delta)$ for which the bulk gaps of $h_\pm$ close, and also specify the gapless momentum locations. At the CDI-NI critical point, $h_\pm$ are both gapless at four Dirac points located at ${\bf k}_{\pm\pm}=(\pm\pi/2,\pm\pi/2)$. For finite $\Delta$, $h_+$ and $h_-$ both take the form of a massive Dirac Hamiltonian in the vicinities of ${\bf k}_{\pm\pm}$. Since $\Delta$ modifies only the mass parameter, $d_{x,y}^+$ and $d_{x,y}^-$ still vanish at ${\bf k}_{\pm\pm}$. However, $d_z^\pm=m\pm|\Delta({\bf k})|+\cos2k_x+\cos2k_y$ will vanish at ${\bf k}_{\pm\pm}$ only for certain values of $(m,\Delta)$. Explicitly, $d_z^+({\bf k}_{\pm\pm})$ vanishes when \begin{equation}
    m+|\Delta({\bf k}_{\pm\pm})|-2=0,\label{eq:h+b}
\end{equation}
and $d_z^-({\bf k}_{\pm\pm})$ vanishes when
\begin{equation}
    m-|\Delta({\bf k}_{\pm\pm})|-2=0.\label{eq:h-b}
\end{equation}
Eqs. \ref{eq:h+b} and \ref{eq:h-b} each define four lines in $(m,\Delta)$ space where the bulk gap of $h_\pm$ closes, i.e., where the reduced Chern number in the appropriate sBZ will change. For example, $m+|\Delta({\bf k}_{++})|-2=0$ describes where $N_r^{++}$ for $h_+$ changes. Since the reduced Chern number of the superconducting system is the sum of the reduced Chern numbers of $h_+$ and $h_-$, the superconducting $\mathcal{N}_r^{++}$ will also change when $m+|\Delta({\bf k}_{++})|-2=0$. 

To determine the phase boundaries we first show that the eight lines defined by Eqs. \ref{eq:h+b} and \ref{eq:h-b} are not all unique. Fermi statistics requires $\Delta({\bf k})$ to be an odd function of momentum since $\tau_x$ is symmetric under orbital exchange. As a result, $|\Delta({\bf k}_{++})|=|\Delta({\bf k}_{--})|$ and $|\Delta({\bf k}_{+-})|=|\Delta({\bf k}_{-+})|$. 
Hence, the putative eight phase boundaries originating from the CDI-NI critical point are reduced to four lines 
\begin{equation}\label{eq:hb}
\begin{split}
m+|\Delta({\bf k}_{++})|-2&=0\\
m+|\Delta({\bf k}_{+-})|-2&=0\\
m-|\Delta({\bf k}_{++})|-2&=0\\
m-|\Delta({\bf k}_{+-})|-2&=0.
\end{split}
\end{equation} 
Because Fermi statistics constrains the gaps in inversion-related (i.e., $C_2$-related in 2D) sBZs to close simultaneously, $C_2$-related sBZs must have the same reduced Chern numbers. The \emph{total} Chern number must therefore always be even for an $n=2$ CDSC, e.g., if the mirror symmetries are preserved as we have been considering, it vanishes. We will see an example below with broken mirror symmetry that has a total Chern number magnitude of two. 

The symmetry of the pairing can further constrain Eq. \ref{eq:hb} and the reduced Chern numbers. If the pairing preserves the mirror symmetries of $h_\pm$, then $|\Delta({\bf k}_{++})|=|\Delta({\bf k}_{+-})|$ and the bulk gap will close in each sBZ simultaneously. In this scenario there are two phase boundaries emanating from the CDI-NI critical point, defining a single intermediate phase characterized by $\mathcal{N}_r^a=-\mathcal{N}_r^b=-1$ (see Fig.~\ref{fig:kx_phase}). In this phase, the reduced Chern numbers of $h_-$ are nontrivial while those of $h_+$ are trivial. Examples of $\Delta({\bf k})$ that preserve the mirror symmetries of $h_\pm$ are $\Delta\sin k_x$ and $\Delta(\sin k_x+i\sin k_y)$; we will consider the former in more detail in the next subsection. 

Interestingly, if the pairing breaks the mirror symmetries of $h_\pm$, the superconducting system can realize phases that have well-defined reduced Chern numbers that add up to nontrivial \emph{total} Chern numbers. A pairing that breaks both mirror symmetries satisfies $|\Delta({\bf k}_{++})|\ne|\Delta({\bf k}_{+-})|$, so the bulk gap does not have to close in each sBZ simultaneously. If $|\Delta({\bf k}_{++})|$ and $|\Delta({\bf k}_{+-})|$ are both nonzero, then all four lines emanating from the CDI-NI critical point in Eq. \ref{eq:hb} are distinct (see Fig.~\ref{fig:kx2ky_phase}). The four lines define three phases, two of which have $\mathcal{N}_r^a\ne-\mathcal{N}_r^b$ and $\mathcal{N}\ne0$. These two phases will be separated by a phase characterized by $\mathcal{N}_r^a=-\mathcal{N}_r^b=-1$. A mirror-breaking pairing that can give rise to the maximal number of phases is $\Delta({\bf k})=\Delta(\sin k_x+c\sin k_y)$ where $|c|\ne1$. For $|c|=1$ the pairing still breaks mirror symmetry but generates fewer phases. Explicitly, if $|c|=1$, then two of the four constraints in Eq. \ref{eq:hb} are identical and correspond to the line $m=2$. At this line, the reduced Chern numbers of both $h_+$ and $h_-$ change. As a result, the system bypasses the $\mathcal{N}_r=\pm1$ phase and jumps between two intermediate $\mathcal{N}\ne0$ phases at $m=2$. In other words, the two lines defining the $\mathcal{N}_r=(-1,1)$ phase in Fig.~\ref{fig:kx2ky_phase} contract to a single line $m=2$ which separates the $\mathcal{N}_r=(-1,2),(-1,0)$ phases. We will study the case $c=2$ in detail below. 

From these general considerations, we see that the $n=2$ CDI and its PH conjugate have the same reduced Chern numbers. As a result, we can realize phases that have nontrivial reduced Chern numbers, and possibly nontrivial total Chern numbers, by placing the $n=2$ CDI in proximity to pairings that preserve or break the mirror symmetries respectively. To confirm this analysis, in the following two subsections we study explicit examples of pairings that preserve or break the mirror symmetries. 

\subsection{Mirror-Preserving Pairing}
\begin{figure}[t]
    \centering
    \includegraphics[width=0.9\linewidth]{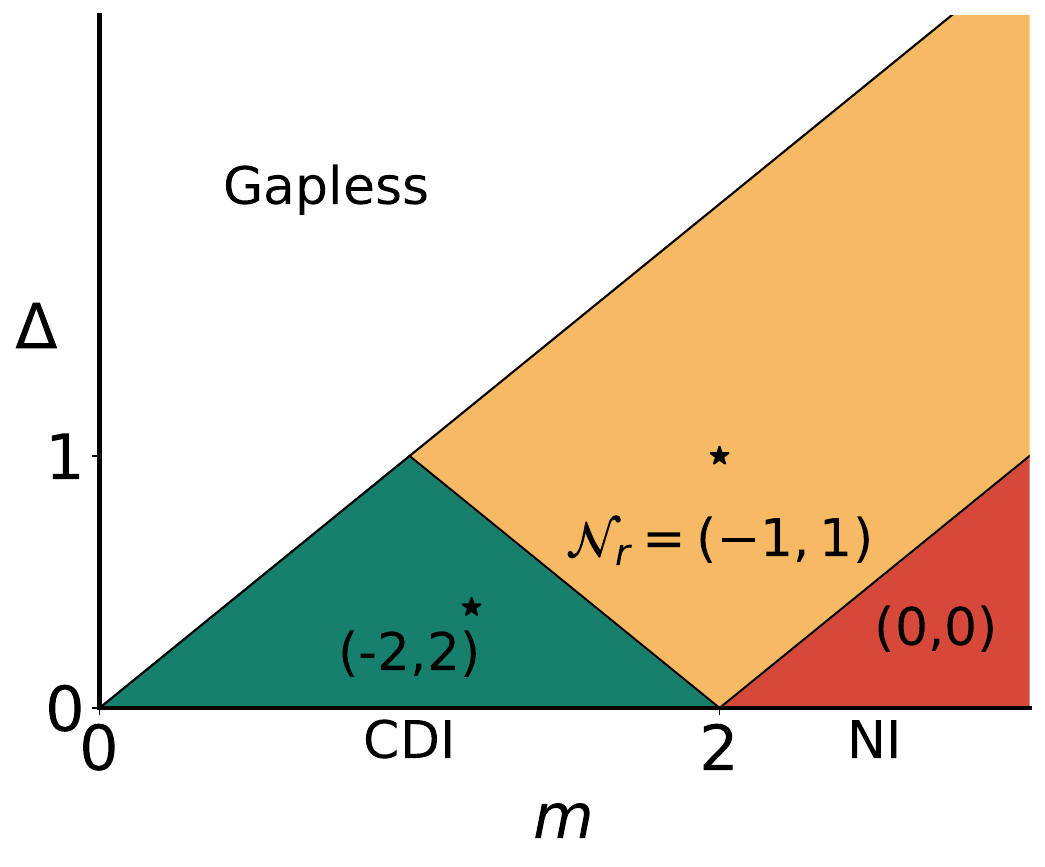}
    \caption{Phase diagram of the CDSC system when a mirror symmetry-preserving $p_x$-wave pairing described by $\Delta({\bf k})=\Delta\sin k_x$ is proximity-induced in the $n=2$ CDI. Pairs  $(\mathcal{N}_r^a,\mathcal{N}_r^b)$ label the reduced Chern numbers for each phase. The CDI and NI phases are well-defined only along $\Delta=0$. The starred points represent parameter values correlated to the edge state spectra shown in Fig.~\ref{fig:kx_edge_spec}a,b. }
    \label{fig:kx_phase}
\end{figure}
\begin{figure}[t]
    \centering
    \includegraphics[width=\linewidth]{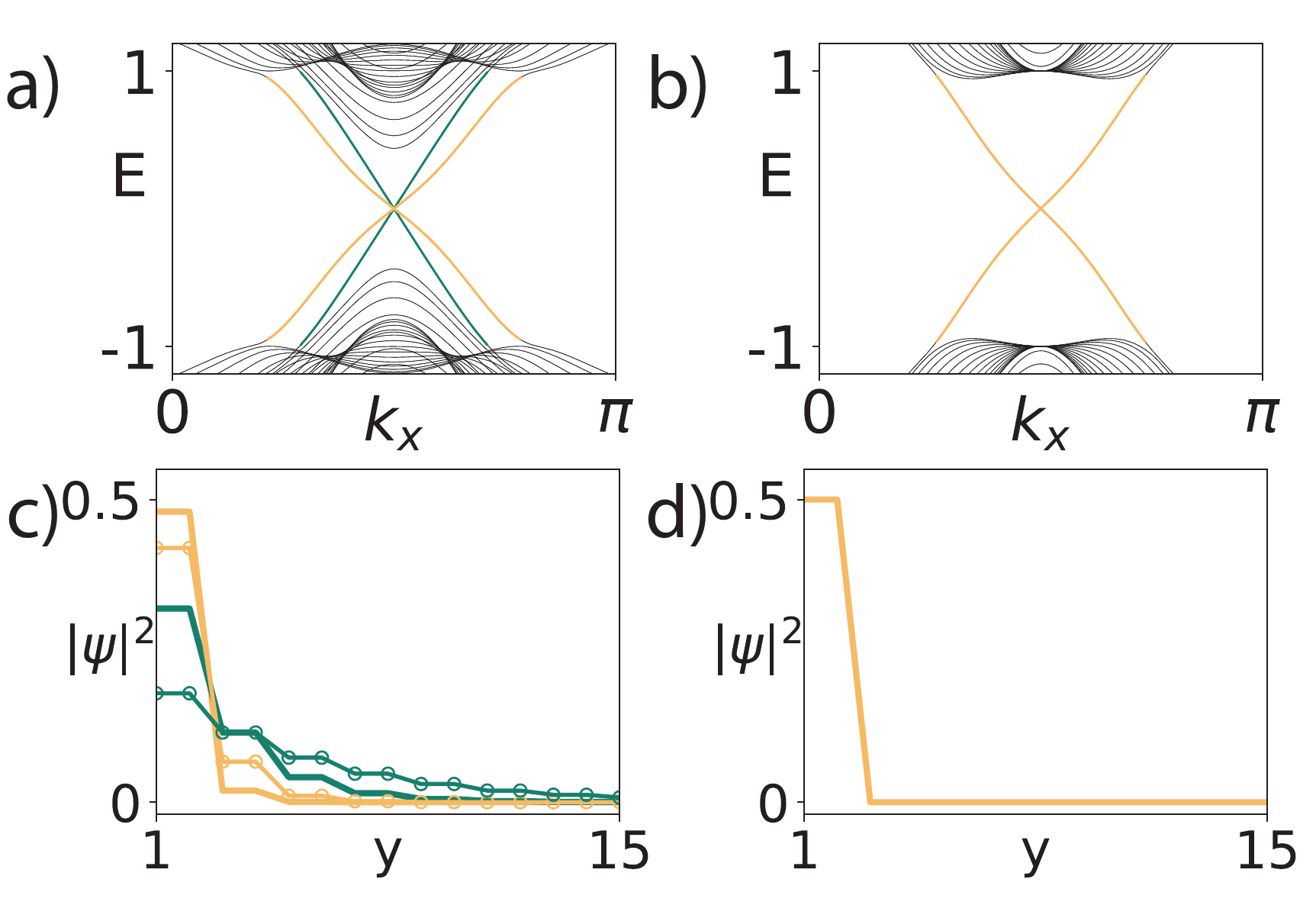}
    \caption{Energy spectra in a cylinder geometry for the $n=2$ CDSC generated by proximitizing a CDI with $p_x$-wave pairing. We have shown only the spectra for $0<k_x<\pi$; the spectra for $-\pi<k_x<0$ are the same due to the mirror symmetry relating the sBZs. (a) and (b) are the spectra at the starred points in the $\mathcal{N}_r=\pm2,\pm1$ phases in Fig.~\ref{fig:kx_phase}, respectively.  The solid lines in (c) and (d) show the real-space probability distribution of the edge states in (a) and (b), respectively, using $L_y=80$. In (c), comparing the solid ($\Delta=0.4$) and circle-marked ($\Delta=0.6$) lines demonstrates that the edge states start to delocalize into the bulk as $\Delta$ increases. 
    }
    \label{fig:kx_edge_spec}
\end{figure}
\begin{figure}[t]
    \centering
    \includegraphics[width=\linewidth]{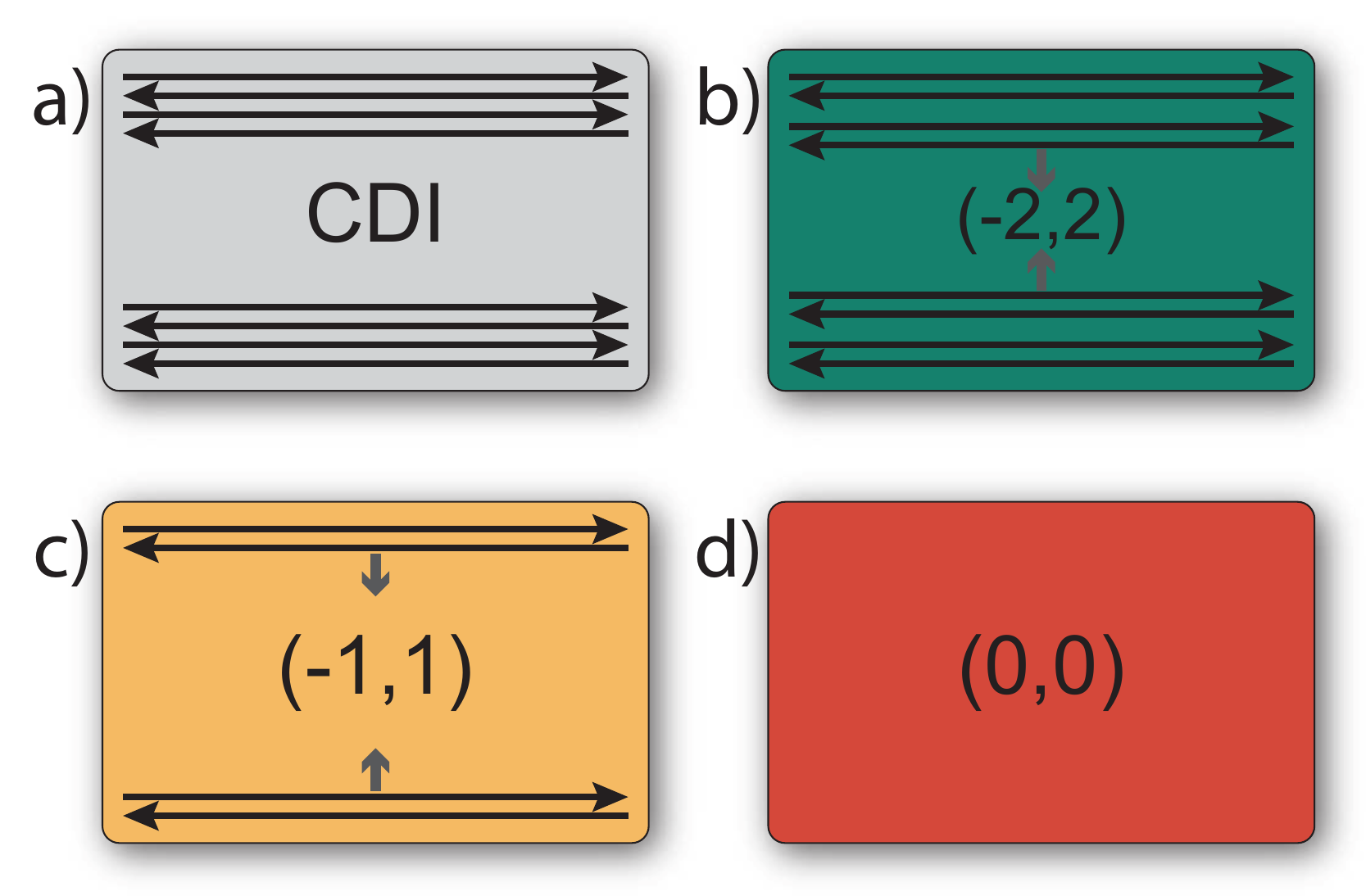}
    \caption{Schematic evolution of the edge states of the $n=2$ CDSC having mirror-preserving pairing. Each edge state line with an arrow represents a chiral Majorana mode in the edge sBZ $0<k_x<\pi$  (an identical picture occurs simultaneously for the sBZ $-\pi<k_x<0$). Pairs $(\mathcal{N}_r^a,\mathcal{N}_r^b)$ label the reduced Chern numbers, and each phase has vanishing total Chern number. (a) The CDI has a pair of complex fermion counter-propagating modes on each edge, which we represent as two pairs of counter-propagating chiral Majorana modes.  Hence, the CDSC in the $\Delta=0$ limit two pairs of counter-propagating chiral Majorana modes on an edge. (b) The pairs become distinct at nonzero $\Delta.$ As $\Delta$ increases, one pair on each edge delocalizes into the bulk and annihilates with the pair from the other edge. (c) The resulting state has a single counter-propagating pair of chiral Majorana modes. (d) Further changes in the parameters induce a phase transition into a trivial SC state that has no Majorana edge modes.}
    \label{fig:kx_edges}
\end{figure}
For our first example we will construct the phase diagram when a mirror symmetry-preserving $p_x$-wave pairing described by $\Delta({\bf k})=\Delta\sin k_x$ ($\Delta\ge0$) is proximity-induced in the $n=2$ CDI. The BdG Hamiltonian has mirror and $C_2$ symmetries represented by $M_x=\mu_0\tau_z$, $M_y=\mu_z\tau_z$, and $C_2=M_xM_y=\mu_z\tau_0$, where the set of $\mu_i$ Pauli matrices acts in the particle-hole space. These symmetries ensure that all gapped phases will have trivial total Chern numbers and quantized, though perhaps vanishing, reduced Chern numbers. 

The BdG Hamiltonian can be block-diagonalized to take the form in Eq. \ref{eq:blockd} where 
\begin{equation}
    h_\pm({\bf k})=h_2({\bf k})\pm\Delta|\sin k_x|\tilde\tau_z.
\end{equation}
Since $|\Delta({\bf k}_{++})|=|\Delta({\bf k}_{+-})|$, there are two phase boundaries in the vicinity of the $(m,\Delta)=(2,0)$ critical point, defined by $\Delta=\mp(m-2)$. When $\Delta<m-2$, both $h_+$ and $h_-$ have trivial reduced Chern numbers. The system is in a trivial $\mathcal{N}_r=0$ superconducting phase that is adiabatically connected to the $N_r=0$ insulator phase in the $\Delta=0$ limit. When $\Delta<-(m-2)$, both $h_+$ and $h_-$ are nontrivial $n=2$ CDIs described by $N_r=\pm1$. The system is then in a nontrivial $\mathcal{N}_r=\pm2$ superconducting phase that is adiabatically connected to the CDI phase in the $\Delta=0$ limit. Between these phases, when $\Delta>|m-2|$, $h_+$ has trivial reduced Chern numbers while $h_-$ has nontrivial reduced Chern numbers. The superconducting system is then in a nontrivial $\mathcal{N}_r=\pm1$ phase. This phase emerges only for finite $\Delta$; it cannot be adiabatically connected to the CDI or NI phases in the $\Delta=0$ limit. The precise phase diagram is shown in Fig.~\ref{fig:kx_phase}.

Just as in the proximitized Chern insulator ($n=0$) case, the phase diagram can be understood through the evolution of the edge states, shown numerically in Fig.~\ref{fig:kx_edge_spec} and schematically in Fig.~\ref{fig:kx_edges}. The states along the $x$ and $y$ edges are qualitatively the same, and here we focus on open boundaries in the $y$-direction where the states are parametrized by $k_x$. Additionally, the edge state spectra for $-\pi<k_x<0$ and $0<k_x<\pi$ are the same because of mirror symmetry relating the sBZs. In what follows we focus on the behavior of the states and spectra for $0<k_x<\pi$ only. 

 For $0<k_x<\pi$ the CDI has a single pair of counter-propagating fermionic modes on a single edge. We can understand these modes as arising from the two sBZ quadrants having opposite reduced Chern numbers that project onto the $0<k_x<\pi$ half of the edge BZ. Thus, the PH-doubled CDSC has two pairs of counter-propagating chiral Majorana modes on a single edge in the $\Delta=0$ limit (Fig.~\ref{fig:kx_edges}a). All four modes, i.e., both pairs, must delocalize and annihilate simultaneously at the CDI-NI critical point. However, finite $\Delta$ lifts the  degeneracy of the Majorana modes. Indeed, the $\mathcal{N}_r=\pm2$ phase at nonzero $\Delta$ has two distinct pairs of degenerate counter-propagating Majorana modes on an edge (Figs.~\ref{fig:kx_edge_spec}a, \ref{fig:kx_edges}b). As $\Delta$ increases, one counter-propagating pair delocalizes into the bulk (Fig.~\ref{fig:kx_edge_spec}c) at a phase transition. After the transition, the resulting $\mathcal{N}_r=\pm1$ state has a single pair of counter-propagating chiral Majorana modes on the edge (Figs.~\ref{fig:kx_edge_spec}b,d, \ref{fig:kx_edges}c).  Since the $\mathcal{N}_r=\pm1$ state has the fewest number of edge modes out of the set of states that have $\mathcal{N}=0$ and $\mathcal{N}_r\ne0$, we say that it is the \emph{minimal} spinless state that has trivial bulk topology and nontrivial sBZ topology. 

\subsection{Mirror-Breaking Pairing}
\begin{figure}
    \centering
    \includegraphics[width=0.9\linewidth]{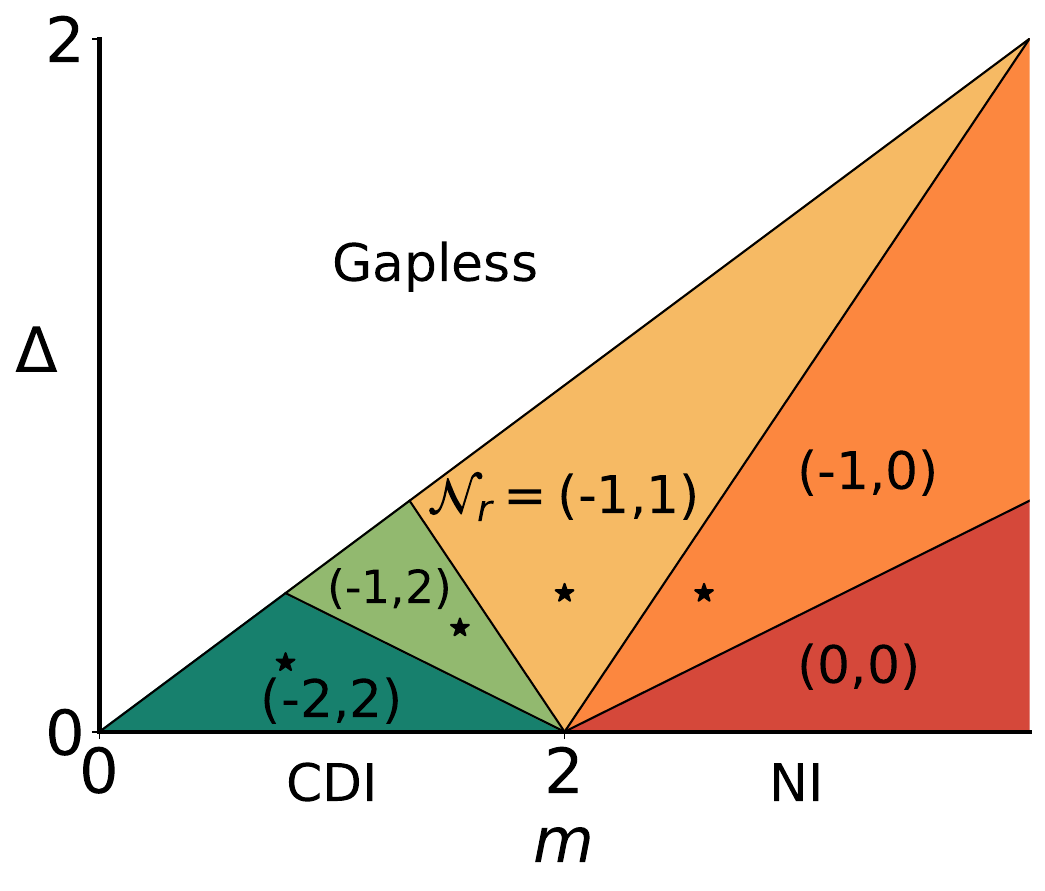}
    \caption{Phase diagram of the CDSC system when a mirror symmetry-breaking pairing described by $\Delta(\sin k_x+2\sin k_y)$ is induced in the $n=2$ CDI. Pairs $(\mathcal{N}_r^a,\mathcal{N}_r^b)$ label the reduced Chern numbers. Starred points represent parameters for the numerical edge state calculations for each phase shown in Fig.~\ref{fig:kx2ky_edge_spec}. }
    \label{fig:kx2ky_phase}
\end{figure}
\begin{figure}
    \centering
    \includegraphics[width=.6\linewidth]{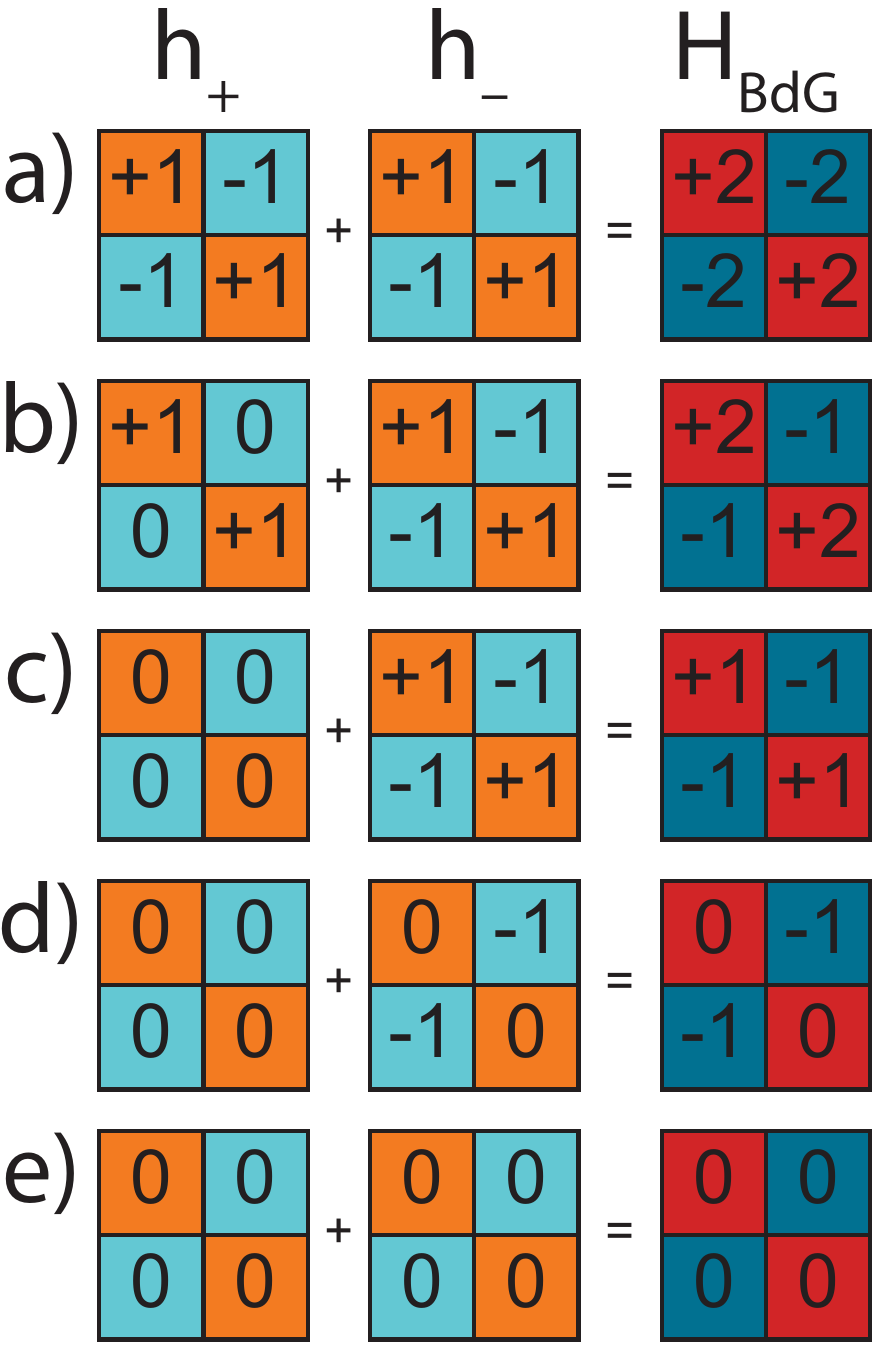}
    \caption{For a mirror symmetry-breaking pairing, the BdG Hamiltonian can be block-diagonalized to the direct sum of two $n=2$ CDIs, $h_+$ and $h_-$, that have mirror symmetry-breaking mass parameters. The reduced Chern numbers of $h_+$ and $h_-$ sum to give the reduced Chern numbers of the superconducting system. (a)-(e) show the reduced Chern numbers for each of the phases in Fig.~\ref{fig:kx2ky_phase} from left to right.
    }
    \label{fig:hp_hm}
\end{figure}
We now study an explicit example of a mirror symmetry-breaking pairing: $\Delta({\bf k})=\Delta(\sin k_x+2\sin k_y)$. This pairing preserves $C_2=\mu_z\tau_0$ symmetry, i.e., the product of the mirror symmetries, and the sBZs related by $C_2$ symmetry must have the same reduced Chern numbers. Crucially, despite the broken mirror symmetries, the reduced Chern numbers remain quantized with this pairing. This is because $h_\pm$ are proportional to $\tilde\tau_z$ along $k_{x,y}=0,\pi$, so the eigenstates are constant, and the Berry curvature still vanishes along the boundaries of each sBZ. Therefore, as mentioned above, the boundary of an sBZ can then be identified as a single point, so the sBZ quadrant is topologically equivalent to $S^2$~\cite{point,dwi}, and hence the reduced Chern numbers are quantized. 

To construct the phase diagram, we block-diagonalize the BdG Hamiltonian to have the form in Eq.~\ref{eq:blockd} where $h\pm$ are defined in Eq.~\ref{eq:hpm2}. The CDSC can then be thought of as the direct sum of two CDI-like systems that have $\Delta$-dependent, mirror-symmetry-breaking mass parameters $m\pm\Delta|\sin k_x+2\sin k_y|$. The phase boundaries in the vicinity of the CDI-NI critical point are defined by $\Delta=\mp(m-2)/3$, where $\mathcal{N}_r^a$ changes, and $\Delta=\mp(m-2)$, where $\mathcal{N}_r^b$ changes. The phase diagram in $(m,\Delta)$ space is shown in Fig.~\ref{fig:kx2ky_phase}, and the reduced Chern numbers of $h_\pm$ and the CDSC for each phase are shown in Fig.~\ref{fig:hp_hm}. 

\begin{figure}
    \centering
    \includegraphics[width=\linewidth]{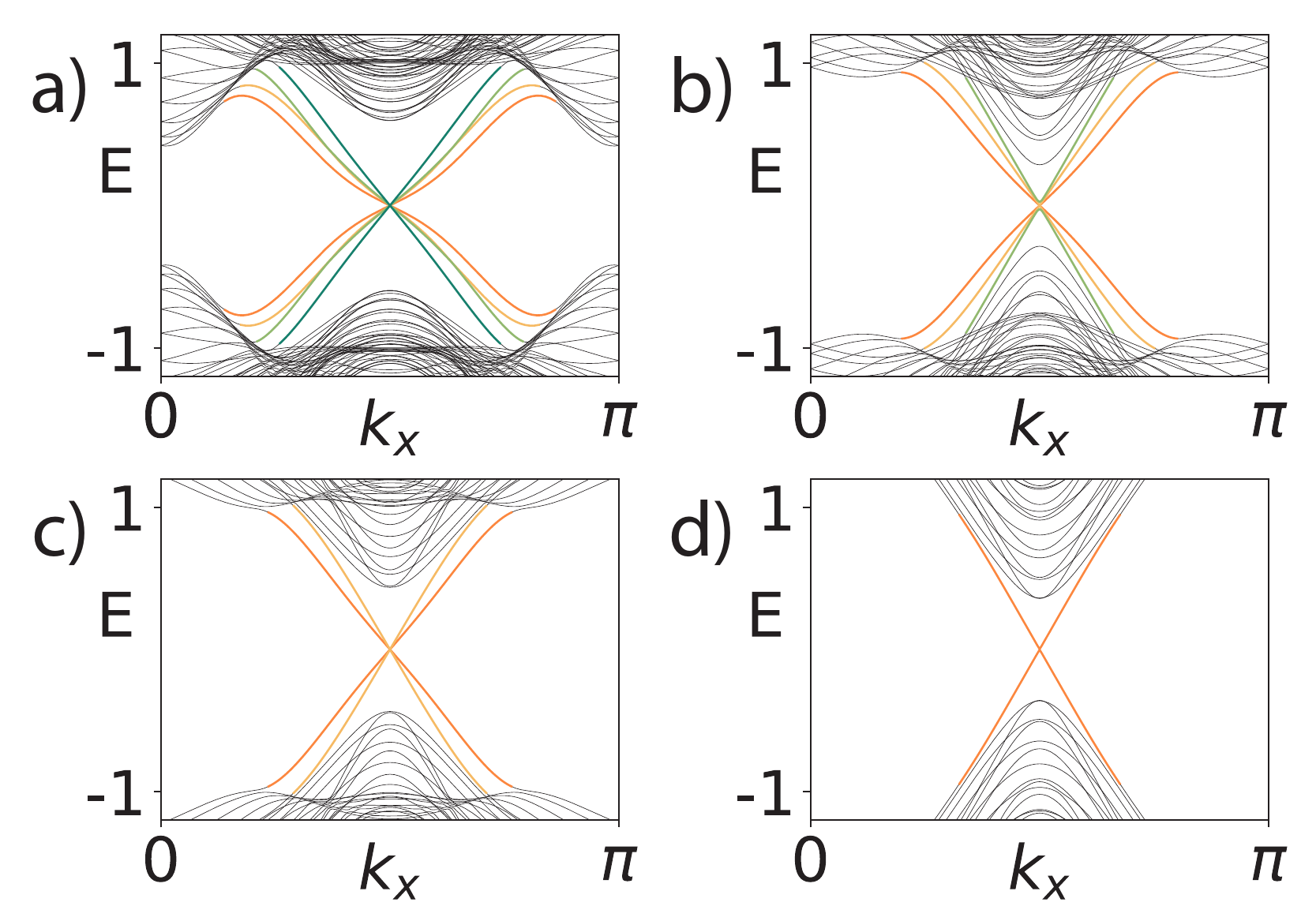}
    \caption{Bulk and edge spectra for the $n=2$ CDSC that has a mirror symmetry-breaking induced pairing described by $\Delta({\bf k})=\Delta(\sin k_x+2\sin k_y)$. Only the spectra for $0<k_x<\pi$ are shown; the spectra for $-\pi<k_x<0$ are the same due to the $C_2$ symmetry relating the sBZs. (a)-(d) correspond to the starred points in the phases of Fig.~\ref{fig:kx2ky_phase} from left to right. The edge states on opposite edges are no longer degenerate due to the broken mirror symmetry. At each phase transition, a single Majorana mode on the edge delocalizes into the bulk, in agreement with how the magnitude of either $\mathcal{N}_r^a$ or $\mathcal{N}_r^b$ decreases by 1.}
    \label{fig:kx2ky_edge_spec}
\end{figure}

\begin{figure}
    \centering
    \includegraphics[width=\linewidth]{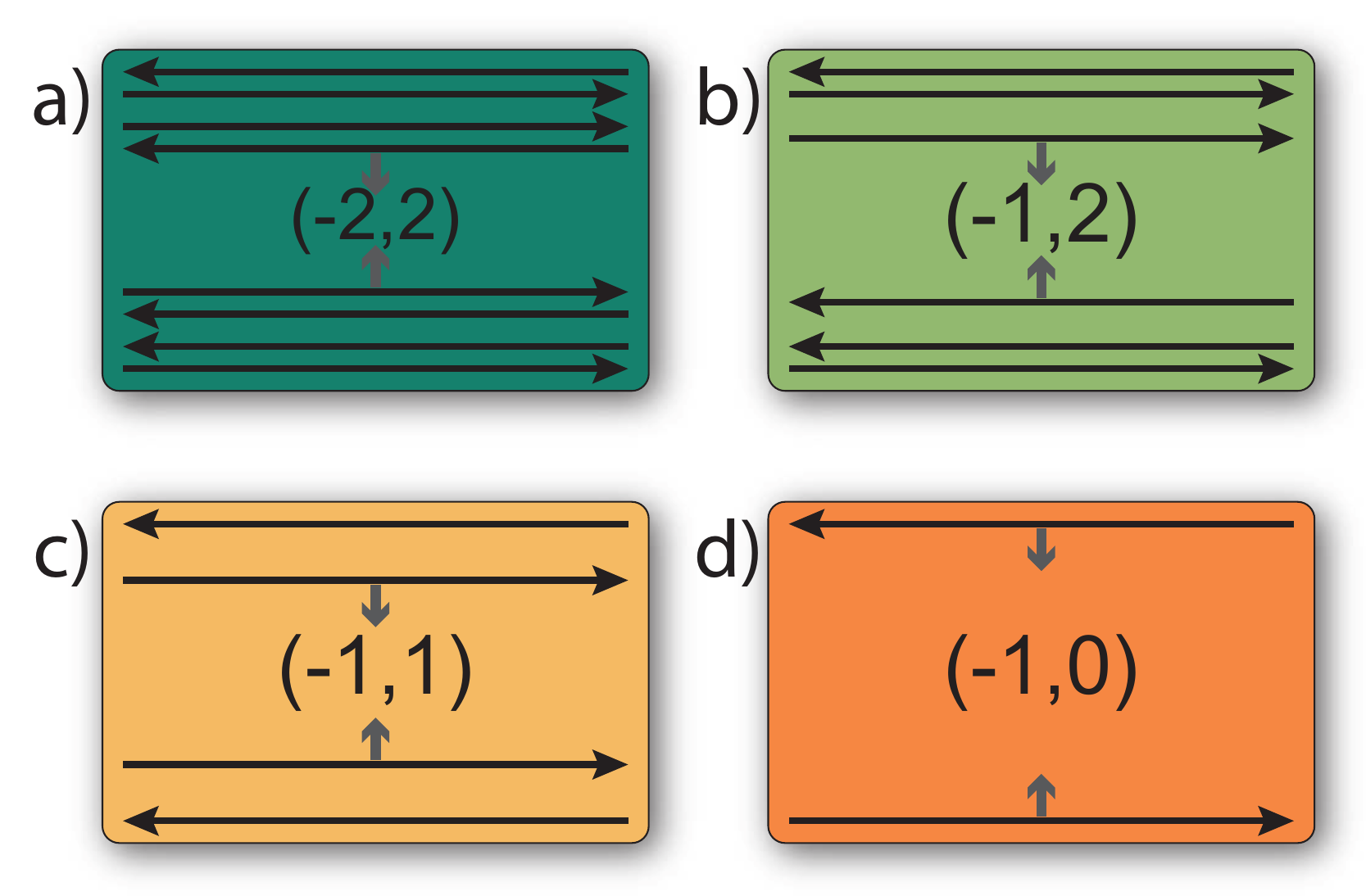}
    \caption{Schematic evolution of the edge states of the $n=2$ CDSC that has a mirror symmetry-breaking pairing. Tuples $(\mathcal{N}_r^a,\mathcal{N}_r^b)$ label the reduced Chern numbers. Each edge state arrow represents a chiral Majorana mode in the edge sBZ $0<k_x<\pi$ (an identical picture occurs simultaneously for the sBZ $-\pi<k_x<0$). At each phase transition, a single Majorana mode on the edge delocalizes into the bulk and annihilates with a mode from the other edge. Each illustration (a)-(d) corresponds to one of the starred points in Fig.~\ref{fig:kx2ky_phase} and one of the numerical spectra in Fig.~\ref{fig:kx2ky_edge_spec}.} 
    \label{fig:kx2ky_edge}
\end{figure}

Just as before, the evolution of the edge states can be used to understand the phase diagram. We again focus on the states and spectra from $0<k_x<\pi$; the behavior of the spectra is identical for $-\pi<k_x<0$. Because the model no longer has mirror symmetry there are two new features. First, the states on mirror-related edges are no longer degenerate with each other. We see this, for example, in Fig.~\ref{fig:kx2ky_edge_spec}a where we can identify four pairs of counter-propagating edge states, where each edge harbors two pairs. The second feature of broken mirror symmetry is that it allows for a net chirality on each edge, i.e., the number of left and right chiral Majorana modes on each edge can be unequal. In the mirror-preserving case, a counter-propagating pair of Majorana modes on each edge annihilated in the bulk at each phase transition, but now we can have cases where only a single chiral Majorana mode from each edge annihilates, as shown schematically in Fig.~\ref{fig:kx2ky_edge}a,b. This can change the net chirality on an edge which is indicated by a change in the total Chern number. 
 
The $n=2$ case exemplifies how an even-$n$ CDI and its PH conjugate have the same reduced Chern numbers. We demonstrated that we can engineer states that have nontrivial reduced and \emph{total} Chern numbers, expanding the classification of phases that have sBZ topology to include strong chiral superconductors. Furthermore, the mirror symmetry-breaking example demonstrates that characterizing phases using sBZ topology and using Berry curvature multipole moments are not always equivalent. That is, phases which have a nonvanishing Chern number, i.e., the monopole moment of Berry curvature, have ill-defined Berry curvature quadrupole moments, whereas we have just demonstrated that a system can have well-defined, quantized reduced Chern numbers that do not sum to zero when considering the entire BZ.

Here, we have studied how odd-momentum $p$-wave pairings generate intermediate phases. However, finding materials with these pairing symmetries may be a challenge. Importantly, one can generate analogous phases by using simpler, even-momentum pairings, e.g., $s$- or extended $s$-wave, at the expense of converting the mirror symmetries of the CDI to momentum-space nonsymmorphic mirror symmetries in the normal-state Hamiltonian. One way of obtaining such a model is to shift the $n=2$ CDI in momentum space by $\pi/2$ in both directions. The shifted $n=2$ CDSC is discussed in more detail in Appendix \ref{sec:n2s}.  


\subsection{Quantized Crystalline Response}
\begin{figure}
    \centering
    \includegraphics[width=0.85\linewidth]{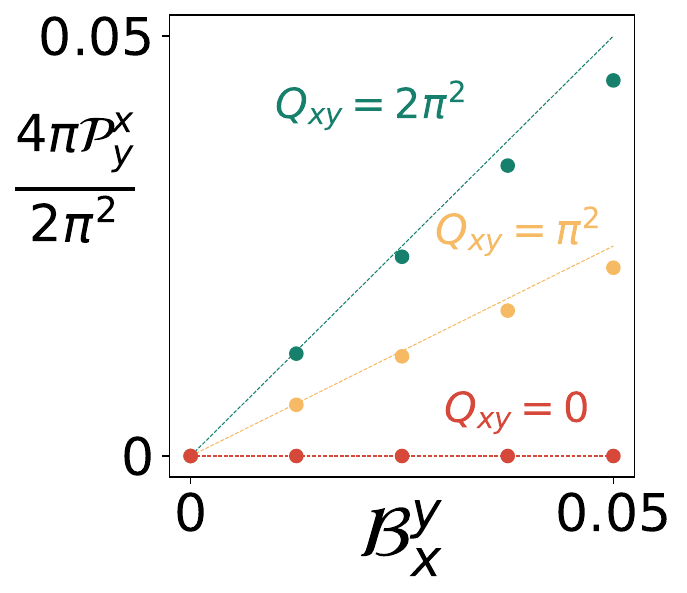}
    \caption{Linear response of the momentum polarization to a lattice shear for the phases of the $n=2$ CDSC that have trivial total Chern numbers. The momentum polarization $\mathcal{P}_y^x$ and translation flux $\mathcal{B}_x^y$ are linearly proportional according to the momentum-weighted Berry curvature quadrupole moment $Q_{xy}$. 
    }
    \label{fig:response}
\end{figure}
We will now show that $n=2$ CDSCs that have a vanishing total Chern number have a nontrivial response encoded in the Berry curvature quadrupole moments. To motivate this we recall that a recent study~\cite{response} has shown that the $n=2$ CDI exhibits a quantized crystalline response in which translational fluxes bind linear momentum charge, i.e., lattice distortions and defects can bind momentum density or generate momentum currents. The response is captured by the effective action~\cite{PhysRevLett.107.075502,emsm},
\begin{equation}
    S_Q=\frac{Q_{ij}}{8\pi}\int\mathfrak{e}^i\wedge d\mathfrak{e}^j,\label{eq:SQ}
\end{equation}
where $\hbar=1$, $Q_{ij}$ is the Berry curvature quadrupole moment defined in Eq. \ref{eq:qxy}, and $\mathfrak{e}^i$ is a background gauge field for translations that encodes strains and dislocations~\cite{nphys1220,PhysRevLett.107.075502,HughesTeo,song2021electric,ThorngrenElse,nissinen2018tetrads,gioia2021unquantized,emsm,PhysRevB.109.075169}. For example, a constant $\mathfrak{e}_x^y$ represents a uniform shear in which electrons moving in the $x$-direction are translated in the $y$-direction; such a gauge field couples to the momentum charge $k_y.$ Additionally, $\nabla\times\mathfrak{e}^i$ represents the dislocation density that has a Burgers vector parallel to the $i$th direction.

Let us focus on $n=2$ CDSCs that have a mirror-preserving pairing so that the total Chern number vanishes and the Berry curvature quadrupole moments are well-defined. Starting in the CDI phase at $\Delta=0$, we know that the moment $Q_{xy}=2\pi^2$ is quantized and the system will exhibit the response in Eq. \ref{eq:SQ}. If we turn on a small $\Delta$ the system remains gapped, and all relevant symmetries are preserved, i.e., translation symmetry so that $\mathfrak{e}^i$ is well-defined, and mirror symmetry so that $Q_{xy}$ is well-defined. Thus, the $n=2$ CDSC phase where $\mathcal{N}_r=\pm 2$ will have the same quantized response as the CDI. This is remarkable because, in many cases, the response of the insulator cannot be applied to the superconductor because the superconductor does not conserve charge~\cite{PhysRevB.103.235427,PhysRevResearch.6.013058,Lutchyn}. For example, the Chern insulator has a well-defined, quantized Hall effect, while the proximitized chiral superconducting phase does not have a quantized charge Hall response since, at least heuristically, we no longer have a well-defined Chern-Simons term built from the electromagnetic gauge fields as soon as $\Delta$ is turned on.

We can confirm the quantized response by direct computation in our lattice model for the $n=2$ CDSC. By varying $S_Q$ with respect to $\mathfrak{e}_y^x$, one can show that there should be a momentum polarization $\mathcal{P}_y^x$ proportional to the quadrupole moment: \begin{equation}
    \mathcal{P}_y^x(t)=\frac{Q_{xy}}{4\pi}\mathfrak{e}_x^y(t).\label{eq:shear}
\end{equation} 
With this specific response term in mind, we can couple the CDSC to such a lattice deformation and see if it exhibits a quantized crystalline response. Explicitly, we can verify the linear relationship in Eq. \ref{eq:shear} by implementing a homogeneous shear via $k_x\rightarrow k_x+\mathcal{B}_x^yk_y$ in the Bloch Hamiltonian. Numerically, for the superconducting case, $\mathcal{P}_y^x$ is calculated by summing states over \emph{half} the BZ to account for the PH redundancy in the BdG formalism. The BdG Hamiltonian artificially doubles the degrees of freedom, so we need to make sure that we are counting only the independent quasiparticle states. Additionally, $Q_{xy}$ is the momentum-weighted Berry curvature quadrupole moment \emph{divided by 2}, such that the $\mathcal{N}_r=\pm2$ CDSC phase has the same quadrupole moment as the $N_r=\pm1$ CDI phase that it is adiabatically connected to in the $\Delta=0$ limit.  

The CDSC should exhibit a quantized crystalline response for the phases that have a well-defined quadrupole moment, i.e., phases that have trivial total Chern numbers. Fig.~\ref{fig:response} shows that the momentum polarization is proportional to the quadrupole moment for the $\mathcal{N}_r=\pm2,\pm1,0$ phases, confirming that the CDSC has the response predicted by the action in Eq. \ref{eq:SQ}. The $\mathcal{N}_r=\pm2$ CDSC state exhibits the same response as the $N_r=\pm1$ CDI state; the CDSC inherits the response of the CDI. The $\mathcal{N}_r=\pm1$ state exhibits half the response of the $\mathcal{N}_r=\pm2$ state, in agreement with the interpretation of the $\mathcal{N}_r=\pm1$ state as half of the $N_r=\pm2$ state. The $\mathcal{N}_r=0$ state exhibits no response of this type.

In summary,  since the response in Eq. \ref{eq:SQ} is independent of electromagnetic gauge fields, the superconductor inherits the well-defined, quantized response from the insulator. This result provides a means of probing the topology of certain phases of the $n=2$ CDSC in experiment. 

\section{Chern dartboard superconductor for \MakeLowercase{n}=1}\label{sec:n1}
\begin{figure}
    \centering
    \includegraphics[width=0.8\linewidth]{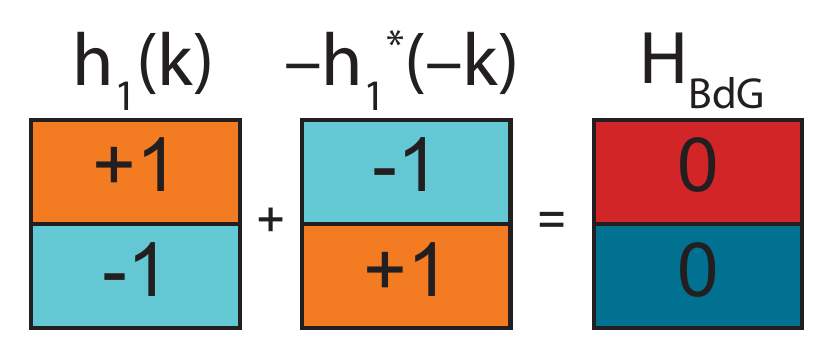}
    \caption{An $n=1$ CDI and its PH conjugate have opposite reduced Chern numbers, so the PH-symmetric direct sum of the two at vanishing superconducting pairing strength has trivial reduced Chern numbers. }
    \label{fig:n1ph}
\end{figure}
\begin{figure}
    \centering
    \includegraphics[width=0.4\linewidth]{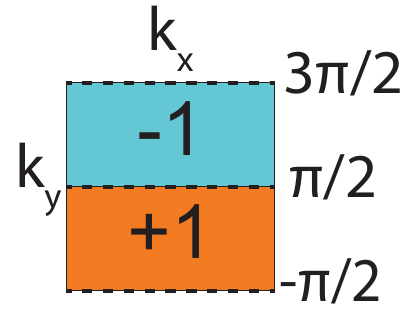}
    \caption{The sBZs for the shifted $n=1$ CDI defined in Eq. \ref{eq:n1CDIshift}. The dashed lines at $k_y=\pm\pi/2$ mark the HSLs of the momentum-space nonsymmorphic mirror symmetry.}
    \label{fig:n1s}
\end{figure}
While the particle and hole copies of even-$n$ CDIs have the same reduced Chern numbers, the particle and hole copies of odd-$n$ CDIs have \emph{opposite} reduced Chern numbers (see Fig.~\ref{fig:n1ph}). As a result, proximitizing an $n=1$ CDI with a superconducting pairing that applies the standard particle-hole transformation will generate a trivial CDSC. However, we will provide two routes to generate a nontrivial $n=1$ CDSC. First, we will show that by changing the sBZ layout of the $n=1$ CDI we can obtain a model which can give rise to nontrivial superconducting reduced Chern numbers. Second, we will mention an alternative approach using an FFLO-type pairing where the Cooper pairs have a finite center-of-mass momentum $Q_y=\pi$~\cite{FF,LO,Casalbuoni_2004}. This second approach modifies the standard PH transformation so that our earlier arguments do not apply.

An $n=1$ CDI has a single mirror symmetry, say $M_y$, which divides the BZ into upper ($0<k_y<\pi$) and lower ($-\pi<k_y<0$) sBZs (see Fig.~\ref{fig:CDI}a). A prototypical model can be constructed as: \begin{multline}
    h_1({\bf k})=\sin k_x\sin k_y\tau_x+\sin2k_y\tau_y\\+(m+\cos k_x+\cos2k_y)\tau_z.\label{eq:n1CDI}
\end{multline}
The mirror symmetry is represented by $M_y=\tau_z,$ where $\tau_i$ acts on an orbital basis, e.g., $s$ and $p_y$ orbitals. For $0<m<2$, the model has reduced Chern numbers $N_r=\pm1$ for the upper and lower sBZs, respectively. This CDI phase can also be characterized by a nontrivial quantized Berry curvature dipole moment $P_y=\pi$, where \begin{equation}
    P_y=\frac{1}{2\pi}\int_{\rm BZ}d^2k\,k_y\mathcal{F}({\bf k}).\label{eq:py}
\end{equation}
The bulk gap closes at $m=0$ and $m=2$. For $|m|>2$, the model is a trivial insulator described by $N_r=P_y=0$, while the system is gapless for $-2\le m\le0$. 

In contrast to the even-$n$ case, an odd-$n$ CDI and its PH conjugate have oppositely-signed $N_r$, so the BdG Hamiltonian without pairing has trivial reduced Chern numbers, as shown schematically in Fig.~\ref{fig:n1ph}. Intuitively, the opposite signs come from how the PH transformation takes ${\bf k}\rightarrow-{\bf k}$, mapping sBZs that have opposite reduced Chern numbers to each other, when under PH they should have the same reduced Chern numbers. We recall that the PH transformation poses no issue for an even-$n$ CDI, where the sBZs at ${\bf k}$ and $-{\bf k}$ have the same reduced Chern numbers. A more detailed discussion of the vanishing reduced Chern numbers for the proximitized $n=1$ case is provided in Appendix \ref{sec:n1pf}. 

While particle-hole symmetry constrains $n=1$ CDSCs to have trivial superconducting reduced Chern numbers, shifting the CDI in momentum space generates a new model which can have nontrivial superconducting sBZ topology at the cost of a more complicated mirror symmetry. Translating the $n=1$ CDI in Eq. \ref{eq:n1CDI} by $\pi/2$ in the $y$ direction gives 
\begin{multline}
    \overline{h}_1({\textbf{k}})=\sin k_x\cos k_y\tau_x-\sin2k_y\tau_y\\+(m+\cos k_x-\cos2k_y)\tau_z.\label{eq:n1CDIshift}
\end{multline}
Although $\overline{h}_1({\textbf{k}})$ lacks mirror symmetry, it has a single momentum-space nonsymmorphic mirror symmetry
\begin{equation}
    \overline{M}_y\overline{h}_1(k_x,k_y){\overline{M}_y}^{-1}=\overline{h}_1(k_x,-k_y+\pi)
\end{equation}
where $\overline{M}_y=\tau_z$. The Hamiltonian also has $C_2$ symmetry represented by $\tau_z$. The sBZs of this model are now bounded by the high-symmetry lines $k_y=\pm\pi/2$, which are symmetric under ${\bf k}\rightarrow-{\bf k}$ (see Fig.~\ref{fig:n1s}). The PH transformation will then map each sBZ to itself, hence avoiding the problem we had in the initial model where the PH transformation related one sBZ to another.

Now let us consider the phases of this model before adding superconductivity. Let $N_r^0$ represent the reduced Chern number for the sBZ from $k_y=-\pi/2$ to $\pi/2$, and let $N_r^\pi$ represent the reduced Chern number for the sBZ from $k_y=\pi/2$ to $k_y=3\pi/2$. For $0<m<2$, this model has $N_r^0=-N_r^\pi=1$, as shown in Fig.~\ref{fig:n1s}. At $m=2$, the gap closes at two Dirac points at ${\bf k}_{1,2}=(\pi,0),(\pi,\pi)$. For $|m|>2$, the system is a trivial insulator described by $N_r^0=N_r^\pi=0$. At $m=0$, the gap closes at the four points ${\bf k}_{3,4,5,6}=(0,0),(0,\pi),(\pi,\pm\pi/2)$. Note that ${\bf k}_{3,4}$ lie within the sBZs while ${\bf k}_{5,6}$ lie on the borders of the sBZs. The bulk spectrum is gapless for $-2\le m\le0$. 

An equivalent route to obtaining nontrivial superconducting sBZ topology is through a Fulde-Ferrell-Larkin-Ovchinnikov (FFLO) state~\cite{FF,LO,Casalbuoni_2004} arising from the unshifted $n=1$ CDI in Eq. \ref{eq:n1CDI}. FFLO states have a modified particle-hole transformation that incorporates a momentum shift, i.e., the Cooper pairs have a nontrivial center-of-mass momentum ${\bf Q}$. The BdG Hamiltonian using the FFLO PH transformation takes the form
\begin{equation}
    H_{\rm FFLO}({\bf k},{\bf Q})=\begin{pmatrix}
    h_1({\bf k}+{\bf Q}/2)&\hat\Delta({\bf k})\\
    \hat\Delta^\dagger({\bf k})&-h_1^*(-{\bf k}+{\bf Q}/2)
    \end{pmatrix}
\end{equation}
where $h_1({\bf{k}})$ is the Bloch Hamiltonian for the $n=1$ CDI from Eq.~\ref{eq:n1CDI} and the Nambu basis is 
 $\begin{pmatrix}c_{{\bf k}+{\bf Q}/2,s}&c_{{\bf k}+{\bf Q}/2,p_y}&c_{-{\bf k}+{\bf Q}/2,s}^\dagger&c_{-{\bf k}+{\bf Q}/2,p_y}^\dagger
\end{pmatrix}^T$. If ${\bf Q}=\pi\hat y$ then the FFLO state proximitizing Eq. \ref{eq:n1CDI} can have nontrivial superconducting sBZ topology. The FFLO approach demonstrates the nontrivial impact of the particle-hole transformation on superconducting sBZ topology. Both approaches---shifting the CDI or using an FFLO PH transformation---give the same BdG Hamiltonian, and for what follows we focus on the shifted-CDI approach. 

When a $\Delta({\bf k})\tau_i$ superconducting pairing is induced, the shifted CDSC can be described by the BdG Hamiltonian 
\begin{equation}
    H_{\rm BdG}=\frac{1}{2}\sum_{\bf k}\Psi_{\bf k}^\dagger\begin{pmatrix}\overline{h}_1({\bf k})&\Delta({\bf k})\tau_i\\\Delta^\dagger({\bf k})\tau_i&-\overline{h}_1^*(-{\bf k})\end{pmatrix}\Psi_{\bf k},\label{eq:hbdg1}
\end{equation}
where $\Psi_{\bf k}=\begin{pmatrix}c_{{\bf k}s}&c_{{\bf k}p_y}&c_{-{\bf k}s}^\dagger&c_{-{\bf k}p_y}^\dagger\end{pmatrix}^T$. In the $\Delta=0$ limit, the BdG Hamiltonian is a direct sum of the shifted CDI Hamiltonian in Eq. \ref{eq:n1CDIshift} and its PH conjugate. Since the reduced Chern numbers for the particle and hole states are equal for this model, the superconducting reduced Chern number $\mathcal{N}_r=N_{p,r}+N_{h,r}$ is twice the reduced Chern number of the shifted CDI: $\mathcal{N}_r=\pm2$ when $0<m<2$ and $\mathcal{N}_r=0$ when $m>2$. 

To study the superconducting system for nonvanishing $\Delta$, the BdG Hamiltonian can be block-diagonalized. Using the matrices in Appendix \ref{sec:n1basis}, the BdG Hamiltonian takes the form in Eq. \ref{eq:blockd} where 
\begin{equation}
    h_\pm({\bf k})=\overline{h}_1({\bf k})+|\Delta({\bf k})|\cdot
    \begin{cases}
    +\tilde\tau_0,&i=x\\
    +\tilde\tau_x,&i=0\\
    +\tilde\tau_y,&i=z\\
    \pm\tilde\tau_z,&i=y\\
    \end{cases}.
\end{equation} The Pauli matrices $\tilde\tau_i$ represent the transformed basis after the block diagonalization, and $h_{\pm}({\bf{k}})$ are the same for cases without the $\pm$ signs. In terms of the transformed basis, $\overline{M}_y=\tilde\tau_z$. The BdG Hamiltonian is thus equivalent to a direct sum of two shifted $n=1$ CDIs in Eq. \ref{eq:n1CDIshift} with $\Delta$-dependent coefficients. Just as in the $n=2$ case, we focus on a $\tilde\tau_z$ pairing that modifies the mass parameters of $h_\pm$. 

The Pauli exclusion principle provides a way to understand why $\tau_{0,x,z}$ pairings cannot generate an intermediate phase. Fermi statistics requires $\Delta({\bf k})$ to be even for a $\tau_y$ pairing and odd for $\tau_{0,x,z}$ matrix structures (these matrices are in the basis before the transformation is applied). The pairing can open a gap at ${\bf k}$ only if $\Delta({\bf k})$ is nonzero. However, if $\Delta({\bf k})$ is odd, then it must be zero at the ${\bf k}_{1,2,3,4}$ values listed above. Therefore, $\tau_{0,x,z}$ pairings cannot give rise to an intermediate phase. 

Unlike the $n=2$ case, where all intermediate phases emerged from the CDI-NI critical point at $(m,\Delta)=(2,0)$, the shifted $n=1$ case can also have intermediate phases originating from the  CDI-gapless critical point at $(m,\Delta)=(0,0)$. For the $n=2$ CDI, the gap closes at $m=0$ only along the borders of the sBZs. However, for the shifted $n=1$ CDI, the gap closes at $m=0$ at points within the sBZs as well. The superconducting pairing can therefore open the gap at these points within the sBZs, giving rise to gapped phases emerging from the CDI-gapless critical point. 

We now consider how the momentum dependence of $\Delta({\bf k})$ influences the reduced Chern numbers of the superconducting system. We first focus on the phase boundaries emerging from the CDI-NI critical point $(m,\Delta)=(2,0)$. At this critical point, $h_\pm$ are gapless at two Dirac points at ${\bf k}_{1,2}=(\pi,0),(\pi,\pi)$. For finite $\Delta$, $d_z^\pm=0$ at ${\bf k}_{1,2}$ when \begin{equation}
    m\pm|\Delta({\bf k}_{1,2})|-2=0.\label{eq:n1dzm2}
\end{equation}
Eq. \ref{eq:n1dzm2} defines four lines in $(m,\Delta)$ space where a superconducting reduced Chern number will change. 

We now construct the phase boundaries emerging from the CDI-gapless critical point $(m,\Delta)=(0,0)$. At this critical point, $h_\pm$ are gapless at ${\bf k}_{3,4,5,6}=(0,0),(0,\pi),(\pi,\pm\pi/2)$. Since ${\bf k}_{3,4}=(0,0),(0,\pi)$ each lie within an sBZ, the closing of the gap at ${\bf k}_{3,4}$ can indicate a change in the reduced Chern number for that sBZ. On the other hand, ${\bf k}_{5,6}=(\pi,\pm\pi/2)$ lie on the borders of the sBZs and will not change the reduced Chern numbers as the gaps are opened and closed. Hence we can focus on just ${\bf k}_{3,4}.$

To determine the phase boundaries across which the reduced Chern numbers change, we note that at nonzero $\Delta$, $d_{x,y}^\pm$ vanish at ${\bf k}_{3,4}$, while $d_z^\pm=0$ at ${\bf k}_{3,4}$ only when 
\begin{equation}
    m\pm|\Delta({\bf k}_{3,4})|=0.
\end{equation}
For $m<0$ the superconducting system is gapless or has trivial reduced Chern numbers, so we focus on $m>0$. Then the two relevant phase boundaries that emerge from $(m,\Delta)=(0,0)$ correspond to when the bulk gap of $h_-$ closes. Explicitly, the two equations for these phase boundaries are 
\begin{equation}
m-|\Delta({\bf k}_{3,4})|=0.\label{eq:n1dzm0}
\end{equation}
While $m-|\Delta({\bf k}_3)|=0$ separates gapped phases that have different reduced Chern numbers, $m-|\Delta({\bf k}_4)|=0$ defines a (bulk) gapless-gapped phase boundary. Altogether, Eqs. \ref{eq:n1dzm2} and \ref{eq:n1dzm0} define lines that generally intersect each other. The system can then pass through a multitude of phases between the gapless, $\mathcal{N}_r=\pm2$, and $\mathcal{N}_r=0$ phases. We will see an explicit example below (see Fig.~\ref{fig:phase1break}).

Analogous to the $n=2$ case, the superconducting pairings can be organized by whether or not they preserve the nonsymmorphic mirror symmetry of $h_\pm$. If the pairing preserves the symmetry, then $|\Delta({\bf k}_{1,2,3,4})|$ are all equivalent and the bulk gap closes in each sBZ simultaneously. In this case there is a single phase boundary originating from the $m=0$ critical point that separates the gapped and gapless phases. There are also two phase boundaries originating from the $m=2$ critical point, defining a single intermediate phase characterized by $\mathcal{N}_r^0=-\mathcal{N}_r^\pi=+1$. Examples of $\Delta({\bf k})$ that preserve the nonsymmorphic mirror symmetry are $\Delta$, $\Delta\cos k_y$, and $\Delta(\cos k_x+i\cos k_y)$. 

If the pairing breaks the nonsymmorphic mirror symmetry of $h_\pm$, the superconducting system can realize phases that have nontrivial total Chern numbers as well. Breaking the symmetry implies $|\Delta({\bf k}_{1,2,3,4})|$ will not all be identical. The maximum number of phases can be obtained when $|\Delta({\bf k}_{1,2,3,4})|$ are all unique and nonvanishing. Since $|\Delta({\bf k}_{1,2,3,4})|$ set the slopes of the phase boundaries in the $(m,\Delta)$ parameter space, unique values allow the phase boundaries to intersect each other. In this scenario, the system can realize a wide variety of gapped phases (see Fig.~\ref{fig:phase1break}). One pairing that generates unique and nonvanishing $|\Delta({\bf k}_{1,2,3,4})|$ is $\Delta({\bf k})=\Delta(c_0+\cos k_x+c_y\cos k_y)$ where $|c_0|,|c_y|\ne1$ and $|c_0|\ne|c_y|$. 
We also find that while some of the phases have the same reduced Chern numbers they can sometimes be distinguished by a $\mathbb{Z}_2$ weak invariant. We will go through an example in more detail below. 

In summary, the PH transformation acts on an odd-$n$ CDI differently than it acts on a Chern insulator or an even-$n$ CDI. An odd-$n$ CDI with mirror symmetry and its PH conjugate have opposite reduced Chern numbers, constraining the odd-$n$ CDSC to have trivial reduced Chern numbers. However, a shifted CDI that has a single momentum-space nonsymmorphic mirror symmetry or a CDI in proximity to an FFLO pairing can give rise to a shifted $n=1$ CDSC that has nontrivial sBZ topology. With different choices of superconducting pairings, this shifted CDSC can realize the ``minimal'' spinless phase that has a trivial total Chern number and the smallest nontrivial $n=1$ reduced Chern numbers, as well as phases that have nontrivial total and reduced Chern numbers. 

\subsection{Mirror-Preserving Pairing}
A momentum-independent $s$-wave pairing is the simplest nonsymmorphic mirror symmetry-preserving pairing that will generate the minimal spinless state that has a trivial total Chern number and nontrivial reduced Chern numbers. Since the pairing has no momentum dependence, the gap opens and closes in each sBZ simultaneously. The phase diagram is the same as that of the $n=2$ mirror symmetry-preserving case as shown in Fig.~\ref{fig:kx_phase}. The edge state evolution is also similar to the $n=2$ case except that whereas the $n=2$ case has pairs of counter-propagating edge states in each edge sBZ, the $n=1$ case has only chiral Majorana modes in each edge sBZ. Focusing on an edge parallel to the $y$-direction and the states with momentum $-\pi/2<k_y<\pi/2$, when $\mathcal{N}_r=\pm2$, there is a pair of co-propagating chiral Majorana modes on each edge. When $\mathcal{N}_r=\pm1$, there is a single chiral Majorana mode on each edge, and there are no Majorana modes when $\mathcal{N}_r=0$.

\subsection{Mirror-Breaking Pairing}
\begin{figure}
    \centering
    \includegraphics[width=\linewidth]{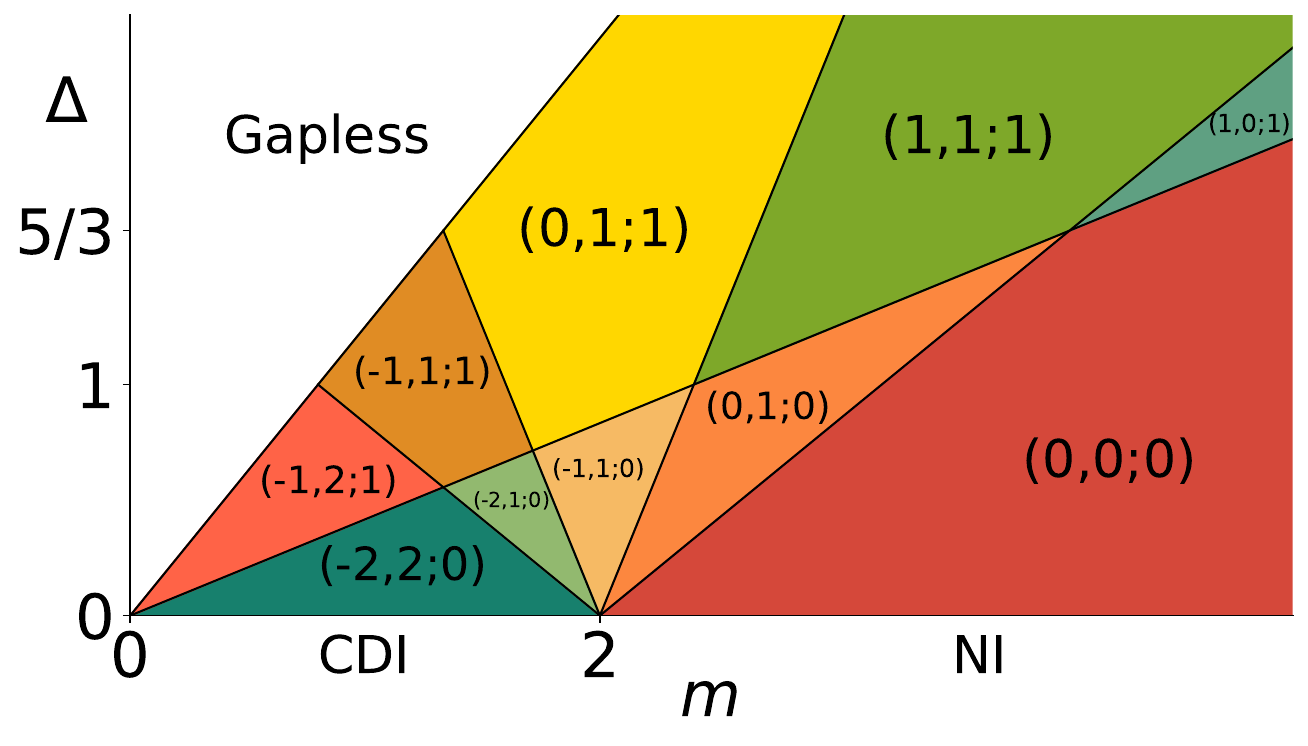}
    \caption{Phase diagram when a mirror symmetry-breaking pairing described by $\Delta(3/5+\cos k_x+(4/5)\cos k_y)$ is induced in the shifted $n=1$ CDI. Phases are labeled by $(\mathcal{N}_r^0,\mathcal{N}_r^\pi;\nu_{x,0})$ where $\nu_{x,0}$ is a weak topological invariant defined in Eq. \ref{eq:weakinv}.}
    \label{fig:phase1break}
\end{figure}
\begin{figure}
    \centering
    \includegraphics[width=.9\linewidth]{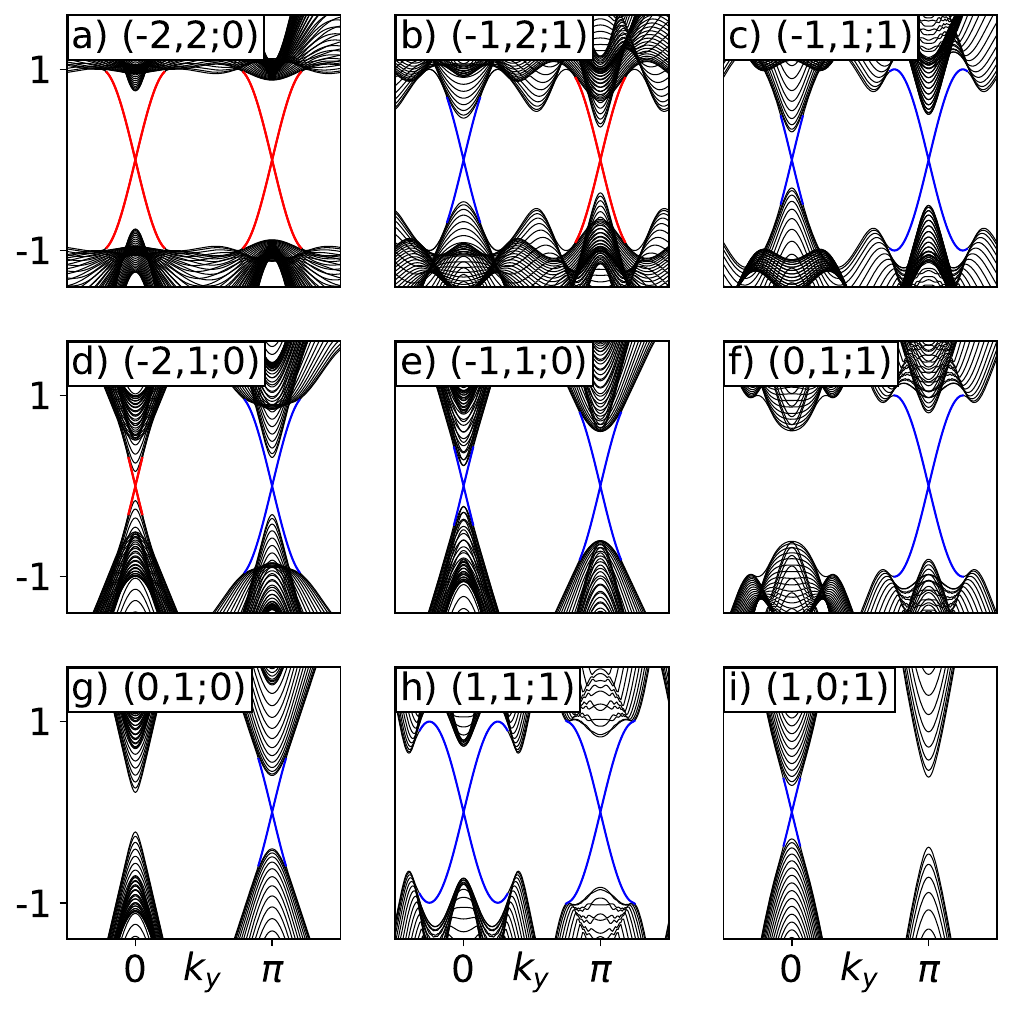}
    \caption{Bulk and edge spectrum as a function of $k_y$ when a pairing described by $\Delta({\bf k})=\Delta(3/5+\cos k_x+(4/5)\cos k_y)$ is induced in the shifted $n=1$ CDI. The labels are $(\mathcal{N}_r^0,\mathcal{N}_r^\pi;\nu_{x,0})$. Disjoint phases which have the same reduced Chern numbers can be distinguished from each other by a weak topological index $\nu_{x,0}$. The magnitudes of $\mathcal{N}_r^{0,\pi}$ can be seen through the degeneracy of the edge states at $k_y=0,\pi$, respectively; red edge states are doubly degenerate while blue edge states are nondegenerate. In order from (a) to (i), the parameters used are $(m,\Delta)=(1.0,0.1)$, $(0.75,0.5)$, $(1.3,1.0)$, $(1.7,0.5)$, $(2.0,0.5)$, $(2.0,1.5)$, $(2.5,0.75)$, $(4.0,3.0)$, $(5.0,2.2)$. 
    }
    \label{fig:n1edgex_break}
\end{figure}
\begin{figure}
    \centering
    \includegraphics[width=.9\linewidth]{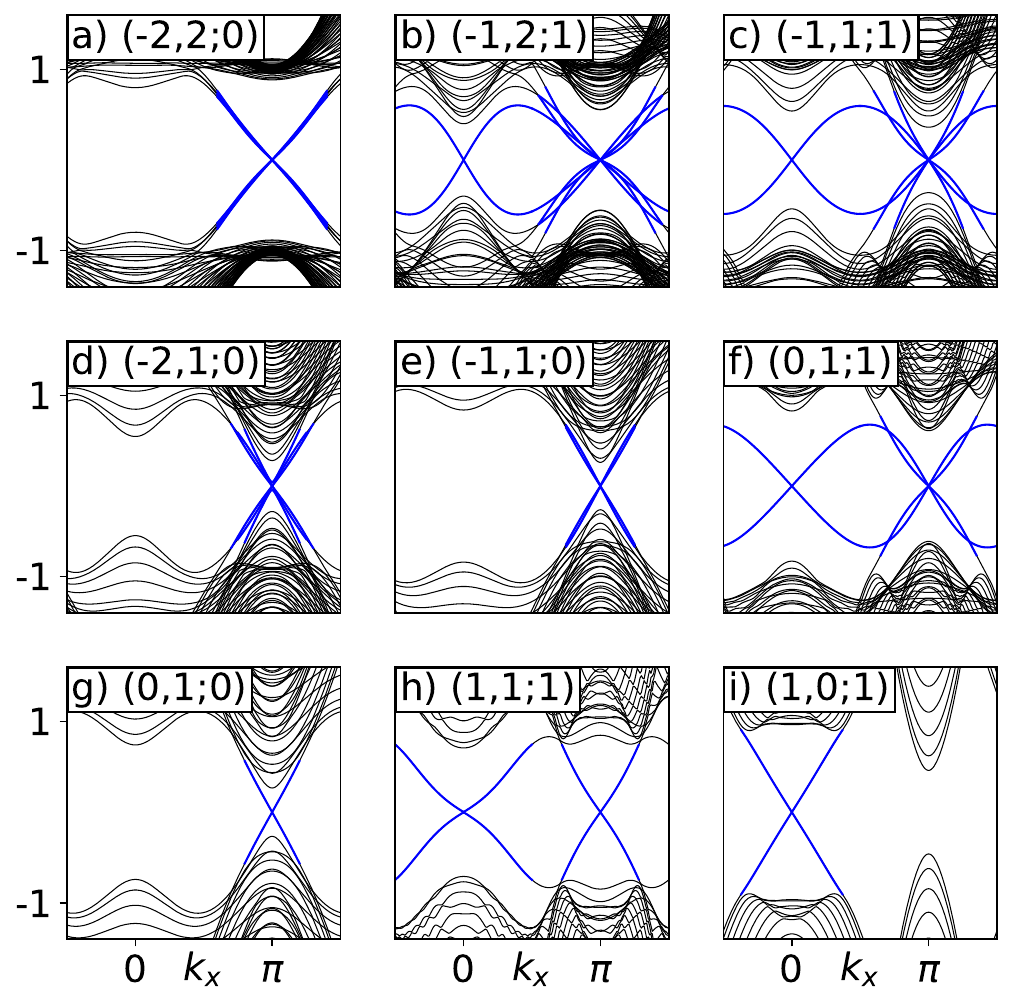}
    \caption{Bulk and edge spectrum as a function of $k_x$ when a pairing described by $\Delta({\bf k})=\Delta(3/5+\cos k_x+(4/5)\cos k_y)$ is induced in the shifted $n=1$ CDI. The labels are $(\mathcal{N}_r^0,\mathcal{N}_r^\pi;\nu_{x,0})$. $\nu_{x,0}$ counts the parity of the number of edge states at $k_x=0$. The plots use the same parameters as Fig.~\ref{fig:n1edgex_break}. 
    }
    \label{fig:n1edgey_break}
\end{figure}
Finally, we construct the phase diagram when $\Delta({\bf k})=\Delta(c_0+\cos k_x+c_y\cos k_y).$ For illustrative purposes we choose $c_0=3/5$ and $c_y=4/5$. Since $h_\pm$ are proportional to $\tilde\tau_z$ when $k_y=\pm\pi/2$, the occupied Bloch eigenstates are constant and the Berry curvature vanishes along the borders of the sBZs of Fig.~\ref{fig:n1s}. The reduced Chern numbers therefore remain quantized despite the broken momentum-space nonsymmorphic mirror symmetry. 

Since $\Delta({\bf k})$ is different at each of the four Dirac points, Eqs. \ref{eq:n1dzm2} and \ref{eq:n1dzm0} define six unique phase boundaries: 
\begin{subequations}
    \begin{align}
        m\pm\frac{2}{5}\Delta-2&=0\label{eq:n1k1}\\
        m\pm\frac{6}{5}\Delta-2&=0\label{eq:n1k2}\\
        m-\frac{12}{5}\Delta&=0\label{eq:n1k3}\\
        m-\frac{4}{5}\Delta&=0\label{eq:n1k4}.
    \end{align}\label{eq:n1k}
\end{subequations}
The first four equations describe phase boundaries originating from the CDI-NI critical point $(m,\Delta)=(2,0)$: Eqs. \ref{eq:n1k1} define where $N_r^{\pm,0}$ change, while Eqs. \ref{eq:n1k2} define where $N_r^{\pm,\pi}$ change. The last two equations describe phase boundaries originating from the CDI-gapless critical point $(m,\Delta)=(0,0)$: Eq. \ref{eq:n1k3} defines where $N_r^{-,0}$ changes, while Eq. \ref{eq:n1k4} separates the gapless phase from the gapped phase where $N_r^{-,\pi}$ is nontrivial. Since the lines defined in Eq. \ref{eq:n1k} intersect, there are ten gapped phases for $m,\Delta>0$ as shown in Fig.~\ref{fig:phase1break}. 

We find that phases in this model that have the same reduced Chern numbers can be distinguished from each other by a weak $\mathbb{Z}_2$ Pfaffian topological invariant~\cite{PhysRevB.86.100504,PhysRevB.76.045302}. Heuristically, a nontrivial $\mathcal{N}_r^{0,\pi}$ indicates that there will be gapless edge states at $k_y=0,\pi$, respectively, when there are open boundaries in the $x$-direction. However, $\mathcal{N}_r^{0,\pi}$ say nothing about the momenta $k_x$ of the gapless edge states when there are open boundaries in the $y$-direction. The momenta of these edge states are instead determined by $\nu_{x,k_x}$. Together, the reduced Chern numbers and weak topological invariant pinpoint the momenta of the gapless edge modes, uniquely identifying the phases of the $n=1$ CDSC via their different boundary spectra.  

To define the weak topological invariant, we first calculate the Pfaffian of the BdG Hamiltonian. The Pfaffian is well-defined only for skew-symmetric matrices, so we apply a unitary transformation to the BdG Hamiltonian in Eq. \ref{eq:hbdg1}: $H_{\rm BdG}'=UH_{\rm BdG}U^{-1}$ where $U=\exp(i\pi\mu_x/4)$. Then $H_{\rm BdG}'$ is skew-symmetric at the four Dirac points ${\bf k}_{1,2,3,4}=(\pi,0),(\pi,\pi),(0,0),(0,\pi)$. For example, at ${\bf k}_1$,
\begin{equation}
H'_{\rm BdG}({\bf k}_1)=i\begin{pmatrix}
0 & -\Delta({\bf k}_1) & -(m-2) & 0 \\
\Delta({\bf k}_1) & 0 & 0 & m-2 \\
m-2 & 0 & 0 & \Delta({\bf k}_1) \\
0 & -(m-2) & -\Delta({\bf k}_1) & 0 
\end{pmatrix},
\end{equation} is skew-symmetric. Then we can define \begin{equation}
    s_{{\bf k}_{1,2,3,4}}=\sgn({\rm pf}(H_{\rm BdG}'({\bf k}_{1,2,3,4})).
\end{equation} The Pfaffian of $H'_{\rm BdG}$ at ${\bf k}_1$ is then $\Delta({\bf k}_1)^2-(m-2)^2$, which for our choice of $\Delta({\bf k})$ simplifies to $4\Delta^2/25-(m-2)^2$, and hence $s_{{\bf k}_{1}}=\sgn(4\Delta^2/25-(m-2)^2).$ 
 
We then define weak $\mathbb{Z}_2$ topological invariants $\nu_{x,0}$ and $\nu_{x,\pi}$ as \begin{subequations}
     \begin{align}
    (-1)^{\nu_{x,0}}&=s_{{\bf k}_3}s_{{\bf k}_4}\label{eq:weakinv}\\
    (-1)^{\nu_{x,\pi}}&=s_{{\bf k}_1}s_{{\bf k}_2}.
\end{align}
\end{subequations}
The invariants $\nu_{x,0}$ and $\nu_{x,\pi}$ can be thought of as weak topological invariants because they rely on the translational symmetry of the lattice and characterize the topology of one-dimensional slices of the 2D BZ that have fixed $k_x=0,\pi$, respectively. When they take their nontrivial value they indicate the presence of an odd number of edge states at $k_x=0,\pi$, respectively. For example, in the $(\mathcal{N}_r^0,\mathcal{N}_r^\pi)=(-1,2)$ phase, $s_{{\bf k}_{1,2,4}}=-1$ while $s_{{\bf k}_3}=1$ so $\nu_{x,0}=1$ while $\nu_{x,\pi}=0$. If we compare with the edge states computed on an edge parallel to the $y$-direction shown in Fig.~\ref{fig:n1edgey_break}b, we see an odd number of edge states at $k_x=0$ and an even number at $k_x=\pi.$

The $s_{{\bf k}_i}$, and hence the weak-invariants, change when the Pfaffian passes through zero. For example, setting $s_{{\bf k}_1}$ to zero generates Eqs. \ref{eq:n1k}a. Thus, the sign of the Pfaffian, and  $\nu_{x,\pi}$, will change at this established phase boundary. The signs $s_{{\bf k}_{2,3,4}}$ change at the phase boundaries determined by Eqs. \ref{eq:n1k}b,c,d, respectively. Only one of $\nu_{x,0}$ and $\nu_{x,\pi}$ is necessary to identify a phase, since~\cite{PhysRevB.86.100504} \begin{equation}
    \mathcal{N}=\mathcal{N}_r^0+\mathcal{N}_r^\pi=(\nu_{x,0}+\nu_{x,\pi})\mod 2.
\end{equation}
Without loss of generality, we choose $\nu_{x,0}$ to distinguish phases that have the same reduced Chern numbers.  

The values of the reduced Chern numbers and weak topological invariant can be understood through the evolution of the edge states parametrized by $k_y$ and $k_x$, respectively. As mentioned above, the magnitudes of the reduced Chern numbers can be seen by introducing an edge parallel to the $y$-direction. Similar to how the superconducting Chern number $\mathcal{N}$ counts the number of chiral Majorana edge modes, the reduced Chern number counts the number of Majorana edge modes per edge sBZ. Fig.~\ref{fig:n1edgex_break} shows two degenerate co-propagating modes on the edge when $|\mathcal{N}_r|=2$, a single chiral edge mode when $|\mathcal{N}_r|=1$, and no gapless edge modes when $\mathcal{N}_r=0$. On the other hand, the value of $\nu_{x,0}$ can be understood by introducing an edge parallel to the $x$-direction where $\nu_{x,0}$ counts the parity of the number of edge modes that have momentum $k_x=0$, as shown in Fig.~\ref{fig:n1edgey_break}. For example, Fig.~\ref{fig:n1edgey_break}a shows no edge modes at $k_x=0$ and so $\nu_{x,0}=0$. In comparison, Fig.~\ref{fig:n1edgey_break}b shows a single edge mode at $k_x=0$ and has $\nu_{x,0}=1$. Each subfigure in Fig.~\ref{fig:n1edgey_break} that has an odd (even) number of edge modes at $k_x=0$ is labeled by $\nu_{x,0}=1(0)$. The details of the distinct phases can therefore be understood through the evolution of the edge states if we separately consider the $x$ and $y$ edges.

Given that one can define $\nu_{x,k_x}$ and use it to describe the edge states parallel to the $x$-direction, one might also ask whether $\nu_{y,k_y}$ can be relevant. Indeed, $\nu_{y,0}$ and $\nu_{y,\pi}$ indicate the parity of the reduced Chern numbers $\mathcal{N}_r^0$ and $\mathcal{N}_r^\pi$, respectively. Since $\nu_{y,0}$ and $\nu_{y,\pi}$ are unable to distinguish between a vanishing and even-integer reduced Chern number, a nonzero value indicates that the phase is not adiabatically connected to the CDI model. 

Ultimately, the shifted $n=1$ case is not entirely analogous to the $n=2$ case because there are additional phase boundaries that emerge from the gapped-gapless critical point that generate a richer phase diagram. 

\subsection{Response}
The quantized crystalline response for an $n=1$ CDI is~\cite{response}
\begin{equation}
    S_D=\frac{P_i}{2\pi}\int\mathfrak{e}^i\wedge dA,\label{eq:SD}
\end{equation}
where $e=\hbar=1$ and $P_i$ is the Berry curvature dipole moment, which is defined in Eq. \ref{eq:py} when $i=y$. When varied with respect to $A_0$, for instance, this action predicts that lattice dislocations bind electric charge. While $S_D$ can predict meaningful responses for charge-conserving insulating systems, the responses do not immediately translate over for superconducting systems. Thus, while the $n=2$ CDI and CDSC exhibit a quantized crystalline response described by the same action, the $n=1$ CDI and CDSC cannot exhibit the same quantized response because superconducting quasiparticles do not have a quantized electric charge. We leave an investigation of the response of the $n=1$ CDSC for future work.

\section{Conclusion}\label{sec:conc}
We have explored the sub-Brillouin zone topology and response of Chern dartboard superconductors. The BdG particle-hole transformation acts differently on even-$n$ and odd-$n$ CDIs. Whereas an even-$n$ CDI and its PH conjugate have the same reduced Chern numbers, an odd-$n$ CDI and its PH conjugate have opposite reduced Chern numbers. We identified superconducting pairings that generate states that have nontrivial total and reduced Chern numbers for both $n=2$ and a modified $n=1$ case, expanding the scope of sBZ topology. 

Additionally, we predict that the $n=2$ CDSC inherits the well-defined, quantized crystalline response from the $n=2$ CDI. By inducing superconductivity in the CDI, we have uncovered a type of response that survives the insulator-superconductor transition. Our work motivates further exploration of the interplay between bulk topology, sBZ topology, and quantized response. 

More specifically, one future direction is to examine odd-$n$ CDSCs in more detail. A remaining question is whether the $n=1$ CDSC can exhibit a well-defined, quantized response. Additionally, it could be interesting to try to construct an $n=3$ CDSC that has nontrivial sBZ topology. Mirror-related sBZs do not map to each other under a momentum-space translation, so a momentum shift does not generate nontrivial sBZ topology for the $n=3$ CDSC, unlike for the $n=1$ CDSC that we discussed here.

Another avenue could be to extend sBZ topology to other topological invariants and other symmetries, and see how particle-hole symmetry constrains the sBZ topology. For example, whereas the Chern number is defined for a band, the Euler invariant is defined for a set of bands, so it could be interesting to construct insulators and superconductors that have reduced Euler invariants. Additionally, mirror symmetry divides the BZ into sectors defined by high-symmetry lines, whereas other crystalline symmetries such as rotation symmetry may give rise to high-symmetry points. It is not immediately obvious how high-symmetry points can define sBZs. We could examine the interplay of particle-hole symmetry and sBZ topology in systems that have different crystalline symmetries. These superconducting systems may reveal other examples of responses that survive the topological insulator-superconductor transition. Our results point to one way that sBZ topology can be expanded, and we expect that sBZ topology can be extended in several other directions.

\section*{Acknowledgements}
TLH thanks Sachin Vaidya and Andre Grossi Fonseca for useful discussions. TLH thanks ARO MURI W911NF2020166 for support. RC acknowledges support from the US Office of Naval Research MURI grant N00014-20-1-2325.

\bibliography{refs}
\appendix
\section{Basis transformations for \texorpdfstring{$n=2$}{n=2}}\label{sec:n2basis}
Here we present the matrices used to block-diagonalize the $n=2$ BdG Hamiltonian via \begin{equation}
    H_{\rm BdG}=M^{-1}\tilde H_{\rm BdG}M.
\end{equation}
\begin{widetext}
\noindent For $\hat\Delta=\Delta({\bf k})\tau_y$: 
\begin{equation}
    M=\begin{pmatrix}
 -\frac{d_z}{d_x+i d_y} & \frac{1}{2}-\frac{d_x}{d_x+i d_y} & \frac{1}{2}-\frac{d_x}{d_x+i d_y} & \frac{d_z}{d_x+i d_y} \\
 -\frac{1}{2} & 0 & 0 & \frac{1}{2} \\
 0 & -\frac{1}{2} & \frac{1}{2} & 0 \\
 \frac{1}{2} & 0 & 0 & \frac{1}{2} \end{pmatrix}
\end{equation}
For $\hat\Delta=\Delta({\bf k})\tau_x$:
\tiny\begin{equation}
M=\begin{pmatrix}
 \frac{i \left(\frac{(d_z-\Delta )\sqrt{\left(| \Delta | -d_z\right){}^2+d_x^2+d_y^2}}{\sqrt{d_x^2+d_y^2+\left(\Delta -d_z\right){}^2}} +(d_z-| \Delta |)\right)}{2 \left(d_x+i d_y\right)} & \frac{\left(d_y+i d_x\right) \sqrt{\left(| \Delta | -d_z\right){}^2+d_x^2+d_y^2}}{2 \left(d_x+i d_y\right) \sqrt{d_x^2+d_y^2+\left(\Delta -d_z\right){}^2}} & -\frac{\left(d_x-i d_y\right) \sqrt{\left(| \Delta | -d_z\right){}^2+d_x^2+d_y^2}}{2 \left(d_x+i d_y\right) \sqrt{d_x^2+d_y^2+\left(\Delta -d_z\right){}^2}} & \frac{\left(\frac{(d_z-\Delta )\sqrt{\left(| \Delta | -d_z\right){}^2+d_x^2+d_y^2}}{\sqrt{d_x^2+d_y^2+\left(\Delta -d_z\right){}^2}} +(d_z-| \Delta |)\right)}{2 \left(d_x+i d_y\right)}\\
 \frac{i}{2} & 0 & 0 & \frac{1}{2} \\
 \frac{i \left(\frac{(d_z+\Delta)\sqrt{\left(| \Delta | +d_z\right){}^2+d_x^2+d_y^2}}{\sqrt{d_x^2+d_y^2+\left(\Delta +d_z\right){}^2}} -(d_z+| \Delta |)\right)}{2 \left(d_x-i d_y\right)} 
 & \frac{i \sqrt{\left(| \Delta | +d_z\right){}^2+d_x^2+d_y^2}}{2 \sqrt{d_x^2+d_y^2+\left(d_z+\Delta \right){}^2}} & \frac{\sqrt{\left(| \Delta | +d_z\right){}^2+d_x^2+d_y^2}}{2 \sqrt{d_x^2+d_y^2+\left(d_z+\Delta \right){}^2}} & \frac{\left(\frac{-(d_z+\Delta)\sqrt{\left(| \Delta | +d_z\right){}^2+d_x^2+d_y^2}}{\sqrt{d_x^2+d_y^2+\left(\Delta +d_z\right){}^2}} +(d_z+| \Delta |)\right)}{2 \left(d_x-i d_y\right)}  \\
 -\frac{i}{2} & 0 & 0 & \frac{1}{2}\end{pmatrix}
\end{equation}\normalsize
For $\hat\Delta=\Delta({\bf k})\tau_z$: 
\tiny\begin{equation}
M=\begin{pmatrix}-\frac{i d_z \left(\frac{\sqrt{\left(| \Delta | -d_x\right){}^2+d_y^2+d_z^2}}{\sqrt{(\Delta-d_x)^2+d_y^2+d_z^2}}+1\right)}{2 \left(-| \Delta | +d_x+i d_y\right)} & \frac{i \left(-d_x+i d_y+\Delta \right) \sqrt{\left(| \Delta | -d_x\right){}^2+d_y^2+d_z^2}}{2 \left(-| \Delta | +d_x+i d_y\right) \sqrt{(\Delta-d_x)^2+d_y^2+d_z^2}} & \frac{\left(-d_x+i d_y+\Delta \right) \sqrt{\left(| \Delta | -d_x\right){}^2+d_y^2+d_z^2}}{2 \left(-| \Delta | +d_x+i d_y\right) \sqrt{(\Delta-d_x)^2+d_y^2+d_z^2}} & \frac{d_z \left(\frac{\sqrt{\left(| \Delta | -d_x\right){}^2+d_y^2+d_z^2}}{\sqrt{(\Delta-d_x)^2+d_y^2+d_z^2}}+1\right)}{2 \left(-| \Delta | +d_x+i d_y\right)} \\
 -\frac{i}{2} & 0 & 0 & \frac{1}{2} \\
 \frac{i d_z \left(1-\frac{\sqrt{\left(| \Delta | +d_x\right){}^2+d_y^2+d_z^2}}{\sqrt{(\Delta+d_x)^2+d_y^2+d_z^2}}\right)}{2 \left(| \Delta | +d_x-i d_y\right)} & -\frac{i \left(d_x-i d_y+\Delta \right) \sqrt{\left(| \Delta | +d_x\right){}^2+d_y^2+d_z^2}}{2 \left(| \Delta | +d_x-i d_y\right) \sqrt{(\Delta+d_x)^2+d_y^2+d_z^2}} & \frac{\left(d_x-i d_y+\Delta \right) \sqrt{\left(| \Delta | +d_x\right){}^2+d_y^2+d_z^2}}{2 \left(| \Delta | +d_x-i d_y\right) \sqrt{(\Delta+d_x)^2+d_y^2+d_z^2}} & \frac{d_z \left(1-\frac{\sqrt{\left(| \Delta | +d_x\right){}^2+d_y^2+d_z^2}}{\sqrt{(\Delta+d_x)^2+d_y^2+d_z^2}}\right)}{2 \left(| \Delta | +d_x-i d_y\right)} \\
 \frac{i}{2} & 0 & 0 & \frac{1}{2} \end{pmatrix}
\end{equation}\normalsize
For $\hat\Delta=\Delta({\bf k})\tau_0$: 
\tiny\begin{equation}
M=\begin{pmatrix}
 -\frac{i d_z \left(\frac{\sqrt{\left(| \Delta | -d_y\right){}^2+d_x^2+d_z^2}}{\sqrt{d_x^2+\left(\Delta -d_y\right){}^2+d_z^2}}+1\right)}{2 \left(| \Delta | +i d_x-d_y\right)} & -\frac{\left(d_x+i \left(\Delta -d_y\right)\right) \sqrt{\left(| \Delta | -d_y\right){}^2+d_x^2+d_z^2}}{2 \left(-i | \Delta | +d_x+i d_y\right) \sqrt{d_x^2+\left(\Delta -d_y\right){}^2+d_z^2}} & -\frac{\left(d_x+i \left(\Delta -d_y\right)\right) \sqrt{\left(| \Delta | -d_y\right){}^2+d_x^2+d_z^2}}{2 \left(-i | \Delta | +d_x+i d_y\right) \sqrt{d_x^2+\left(\Delta -d_y\right){}^2+d_z^2}} & \frac{d_z \left(\frac{\sqrt{\left(| \Delta | -d_y\right){}^2+d_x^2+d_z^2}}{\sqrt{d_x^2+\left(\Delta -d_y\right){}^2+d_z^2}}+1\right)}{2 \left(-i | \Delta | +d_x+i d_y\right)} \\
 -\frac{1}{2} & 0 & 0 & \frac{1}{2} \\
 \frac{i d_z \left(1-\frac{\sqrt{\left(| \Delta | +d_y\right){}^2+d_x^2+d_z^2}}{\sqrt{d_x^2+\left(d_y+\Delta \right){}^2+d_z^2}}\right)}{2 \left(| \Delta | +i d_x+d_y\right)} & -\frac{\left(i d_x+d_y+\Delta \right) \sqrt{\left(| \Delta | +d_y\right){}^2+d_x^2+d_z^2}}{2 \left(| \Delta | +i d_x+d_y\right) \sqrt{d_x^2+\left(d_y+\Delta \right){}^2+d_z^2}} & \frac{\left(i d_x+d_y+\Delta \right) \sqrt{\left(| \Delta | +d_y\right){}^2+d_x^2+d_z^2}}{2 \left(| \Delta | +i d_x+d_y\right) \sqrt{d_x^2+\left(d_y+\Delta \right){}^2+d_z^2}} & \frac{i d_z \left(1-\frac{\sqrt{\left(| \Delta | +d_y\right){}^2+d_x^2+d_z^2}}{\sqrt{d_x^2+\left(d_y+\Delta \right){}^2+d_z^2}}\right)}{2 \left(| \Delta | +i d_x+d_y\right)} \\
 \frac{1}{2} & 0 & 0 & \frac{1}{2}
\end{pmatrix}
\end{equation}\normalsize

\section{Basis transformations for shifted \texorpdfstring{$n=1$}{n=1}}\label{sec:n1basis}
Here we present the matrices used to block-diagonalize the shifted $n=1$ BdG Hamiltonian.

For $\hat\Delta=\Delta({\bf k})\tau_0$:
\tiny\begin{equation}
    M=\begin{pmatrix}
        
 \frac{i d_z \left(\frac{\sqrt{\left(| \Delta | -d_x\right){}^2+d_y^2+d_z^2}}{\sqrt{(\Delta-d_x)^2+d_y^2+d_z^2}}+1\right)}{2 \left(-| \Delta | +d_x+i d_y\right)} & \frac{\left(i d_x+d_y-i \Delta \right) \sqrt{\left(| \Delta | -d_x\right){}^2+d_y^2+d_z^2}}{2 \left(-| \Delta | +d_x+i d_y\right) \sqrt{(\Delta-d_x)^2+d_y^2+d_z^2}} & \frac{\left(d_x-i d_y-\Delta \right) \sqrt{\left(| \Delta | -d_x\right){}^2+d_y^2+d_z^2}}{2 \left(-| \Delta | +d_x+i d_y\right) \sqrt{(\Delta-d_x)^2+d_y^2+d_z^2}} & \frac{d_z \left(\frac{\sqrt{\left(| \Delta | -d_x\right){}^2+d_y^2+d_z^2}}{\sqrt{(\Delta-d_x)^2+d_y^2+d_z^2}}+1\right)}{2 \left(-| \Delta | +d_x+i d_y\right)} \\
 \frac{i}{2} & 0 & 0 & \frac{1}{2} \\
 \frac{i d_z \left(1-\frac{\sqrt{\left(| \Delta | +d_x\right){}^2+d_y^2+d_z^2}}{\sqrt{(\Delta+d_x)^2+d_y^2+d_z^2}}\right)}{2 \left(| \Delta | +d_x-i d_y\right)} & -\frac{i \left(d_x-i d_y+\Delta \right) \sqrt{\left(| \Delta | +d_x\right){}^2+d_y^2+d_z^2}}{2 \left(| \Delta | +d_x-i d_y\right) \sqrt{(\Delta+d_x)^2+d_y^2+d_z^2}} & \frac{\left(d_x-i d_y+\Delta \right) \sqrt{\left(| \Delta | +d_x\right){}^2+d_y^2+d_z^2}}{2 \left(| \Delta | +d_x-i d_y\right) \sqrt{(\Delta+d_x)^2+d_y^2+d_z^2}} & \frac{d_z \left(\frac{\sqrt{\left(| \Delta | +d_x\right){}^2+d_y^2+d_z^2}}{\sqrt{(\Delta+d_x)^2+d_y^2+d_z^2}}-1\right)}{2 \left(| \Delta | +d_x-i d_y\right)} \\
 -\frac{i}{2} & 0 & 0 & \frac{1}{2} 
    \end{pmatrix}
\end{equation}\normalsize
For $\hat\Delta=\Delta({\bf k})\tau_x$:
\begin{equation}
    M=\begin{pmatrix}
        
 \frac{i d_z}{d_x+i d_y} & \frac{d_x}{d_y-i d_x}-\frac{i}{2} & -\frac{1}{2}+\frac{d_x}{d_x+i d_y} & \frac{d_z}{d_x+i d_y} \\
 \frac{i}{2} & 0 & 0 & \frac{1}{2} \\
 0 & -\frac{i}{2} & \frac{1}{2} & 0 \\
 -\frac{i}{2} & 0 & 0 & \frac{1}{2}
    \end{pmatrix}
\end{equation}
For $\hat\Delta=\Delta({\bf k})\tau_y$:
\tiny\begin{equation}
    M=\begin{pmatrix}
        \frac{\frac{(\Delta-d_z)  \sqrt{\left(| \Delta | -d_z\right){}^2+d_x^2+d_y^2}}{\sqrt{d_x^2+d_y^2+\left(\Delta -d_z\right){}^2}}+(| \Delta |-d_z) }{2 \left(d_x+i d_y\right)}& -\frac{\left(d_x-i d_y\right) \sqrt{\left(| \Delta | -d_z\right){}^2+d_x^2+d_y^2}}{2 \left(d_x+i d_y\right) \sqrt{d_x^2+d_y^2+\left(\Delta -d_z\right){}^2}} & \frac{\left(d_x-i d_y\right) \sqrt{\left(| \Delta | -d_z\right){}^2+d_x^2+d_y^2}}{2 \left(d_x+i d_y\right) \sqrt{d_x^2+d_y^2+\left(\Delta -d_z\right){}^2}} & -\frac{\frac{(\Delta-d_z)  \sqrt{\left(| \Delta | -d_z\right){}^2+d_x^2+d_y^2}}{\sqrt{d_x^2+d_y^2+\left(\Delta -d_z\right){}^2}}+(| \Delta |-d_z) }{2 \left(d_x+i d_y\right)}  \\
 -\frac{1}{2} & 0 & 0 & \frac{1}{2} \\
\frac{\frac{(\Delta+d_z)  \sqrt{\left(| \Delta | +d_z\right){}^2+d_x^2+d_y^2}}{\sqrt{d_x^2+d_y^2+\left(d_z+\Delta \right){}^2}}-(| \Delta |+d_z) }{2 \left(d_x-i d_y\right)}& \frac{\sqrt{\left(| \Delta | +d_z\right){}^2+d_x^2+d_y^2}}{2 \sqrt{d_x^2+d_y^2+\left(d_z+\Delta \right){}^2}} & \frac{\sqrt{\left(| \Delta | +d_z\right){}^2+d_x^2+d_y^2}}{2 \sqrt{d_x^2+d_y^2+\left(d_z+\Delta \right){}^2}} &\frac{\frac{(\Delta+d_z)  \sqrt{\left(| \Delta | +d_z\right){}^2+d_x^2+d_y^2}}{\sqrt{d_x^2+d_y^2+\left(d_z+\Delta \right){}^2}}-(| \Delta |+d_z) }{2 \left(d_x-i d_y\right)}\\
 \frac{1}{2} & 0 & 0 & \frac{1}{2}
    \end{pmatrix}
\end{equation}\normalsize
For $\hat\Delta=\Delta({\bf k})\tau_z$:
\tiny\begin{equation}
    M=\begin{pmatrix}
        \frac{d_z \left(\frac{\sqrt{\left(| \Delta | -d_y\right){}^2+d_x^2+d_z^2}}{\sqrt{d_x^2+\left(\Delta -d_y\right){}^2+d_z^2}}+1\right)}{2 \left(-i | \Delta | +d_x+i d_y\right)} & \frac{\left(d_x+i \left(\Delta -d_y\right)\right) \sqrt{\left(| \Delta | -d_y\right){}^2+d_x^2+d_z^2}}{2 \left(-i | \Delta | +d_x+i d_y\right) \sqrt{d_x^2+\left(\Delta -d_y\right){}^2+d_z^2}} & \frac{\left(d_x+i \left(\Delta -d_y\right)\right) \sqrt{\left(| \Delta | -d_y\right){}^2+d_x^2+d_z^2}}{2 \left(-i | \Delta | +d_x+i d_y\right) \sqrt{d_x^2+\left(\Delta -d_y\right){}^2+d_z^2}} & \frac{d_z \left(\frac{\sqrt{\left(| \Delta | -d_y\right){}^2+d_x^2+d_z^2}}{\sqrt{d_x^2+\left(\Delta -d_y\right){}^2+d_z^2}}+1\right)}{2 \left(-i | \Delta | +d_x+i d_y\right)} \\
 \frac{1}{2} & 0 & 0 & \frac{1}{2} \\
 \frac{i d_z \left(1-\frac{\sqrt{\left(| \Delta | +d_y\right){}^2+d_x^2+d_z^2}}{\sqrt{d_x^2+\left(d_y+\Delta \right){}^2+d_z^2}}\right)}{2 \left(|\Delta | +i d_x+d_y\right)} & -\frac{\left(i d_x+d_y+\Delta \right) \sqrt{\left(| \Delta | +d_y\right){}^2+d_x^2+d_z^2}}{2 \left(| \Delta | +i d_x+d_y\right) \sqrt{d_x^2+\left(d_y+\Delta \right){}^2+d_z^2}} & \frac{\left(i d_x+d_y+\Delta \right) \sqrt{\left(| \Delta | +d_y\right){}^2+d_x^2+d_z^2}}{2 \left(| \Delta | +i d_x+d_y\right) \sqrt{d_x^2+\left(d_y+\Delta \right){}^2+d_z^2}} & -\frac{i d_z \left(1-\frac{\sqrt{\left(| \Delta | +d_y\right){}^2+d_x^2+d_z^2}}{\sqrt{d_x^2+\left(d_y+\Delta \right){}^2+d_z^2}}\right)}{2 \left(| \Delta | +i d_x+d_y\right)} \\
 -\frac{1}{2} & 0 & 0 & \frac{1}{2}
    \end{pmatrix}
\end{equation}\normalsize
\end{widetext}

\section{CDSC for \texorpdfstring{$n=1$}{n=1}}\label{sec:n1pf}
In this subsection we will provide proofs that the simplest models for $n=1$ Chern dartboard superconductors necessarily have vanishing reduced Chern numbers. For $n=1$, without loss of generality, let us assume that we have a mirror symmetry $M_y$ that sends $y\to -y.$ Since it is $n=1$ we, by definition, should not have another mirror symmetry such as $M_x,$ and hence we also do not have a $C_2$ symmetry where $C_2=M_x M_y.$ Indeed, if we had an $n=1$ CDI or CDSC, such a $C_2$ symmetry would immediately imply that the reduced Chern numbers vanish. To see this we note that $C_2$ rotates one sector of the BZ to the other and the Chern number is invariant under rotation. Hence, the reduced Chern numbers must have both the same (from $C_2$) and opposite (from mirror) values, and hence must vanish since the Chern number is an \emph{integer} (i.e., not $\mathbb{Z}_2$) invariant.  For example, if we were instead considering a quantized charge polarization, or some other $\mathbb{Z}_2$ quantity in an sBZ, then this would not require that quantity to vanish, just that that quantity is \emph{equivalent} to it negative, i.e., equal to its negative modulo some quantized ambiguity. This leaves the door open to other possible sBZ topological phases for $n=1.$

Now, to prove that a broad class of $n=1$ models cannot support CDSCs we will show that the BdG doubling process introduces a PH symmetry that acts as an effective $C_2$ symmetry. To see this we will focus on two families of models (i) two band models with Bloch Hamiltonians $H({\bf{k}})=\sum_a d_a({\bf{k}})\sigma^a$ where $\sigma^a$ are Pauli matrices, and (ii) four-band Dirac-like models where $H({\bf{k}})=\sum_a d_a({\bf{k}})\Gamma^a$ where $\Gamma^a$ are a set of five $4\times 4$ Dirac matrices.

Let us first consider the two-band models. The normal-state Hamiltonian is $H({\bf{k}})=\sum_a d_a({\bf{k}})\sigma^a.$ This Hamiltonian obeys $-H^{T}({\bf{k}})=\sigma^y H({\bf{k}}) \sigma^y.$ Hence, the particle-hole conjugate (which also requires flipping ${\bf{k}}\to-{\bf{k}}$) is $\sigma^y H(-{\bf{k}})\sigma^y.$ But this is nothing but an effective $C_2$ transformation with representation $C_2=\sigma^y$ (or $i\sigma^y$). Thus, if our normal-state Hamiltonian is in a CDI state for $n=1$, the PH conjugate has opposite reduced Chern numbers. Hence, the PH-symmetric combination has vanishing reduced Chern numbers and has trivial sBZ topology. This is in sharp contrast to the $n=0$ Chern insulator that has a PH conjugate that has the \emph{same} Chern number. Hence, instead of starting from a superconducting system that has twice the (superconducting) Chern number in each sector (as in the $n=0,2$ cases), the $n=1$ system has vanishing reduced Chern number in each sector.

Next, let us consider Dirac-like models of the form $H({\bf{k}})=\sum_a d_a({\bf{k}})\Gamma^a.$ For the five $\Gamma^a$ we can choose the representation $\Gamma^{(1,2,3)}=\tau^x\otimes \sigma^{(x,y,z)}, 
\Gamma^{0}=\tau^z\otimes\mathbb{I}, \Gamma^5=\tau^y\otimes \mathbb{I}.$ We see that two out of the five matrices are imaginary: $\Gamma^2,\Gamma^5$. To construct the PH conjugate we need to consider $-H^{T}({\bf{k}})=(i\Gamma^2\Gamma^5)H({\bf{k}})(i\Gamma^2\Gamma^5).$ Since the PH conjugate also inverts momentum, we see that the BdG charge conjugation is generated by an effective $C_2$ operation $C_2=i\Gamma^2\Gamma^5.$ Hence, as above, we can immediately conclude that models of this type cannot generate an $n=1$ CDSC.

\section{Momentum-shifted \texorpdfstring{$n=2$}{n=2} CDSC}\label{sec:n2s}
In the main text, we argued that the $n=2$ CDSC required a pairing with a symmetric matrix structure to realize intermediate states that were not adiabatically connected to the CDI in the $\Delta=0$ limit. However, these pairings might be more difficult to realize experimentally compared to an $s$-wave pairing, which is asymmetric in the matrix structure. 

If we shift the $n=2$ CDI in momentum space by $k_x=k_y=\pi/2$, converting the two mirror symmetries to momentum-space nonsymmorphic mirror symmetries, then we can obtain intermediate phases using $s$- and extended $s$-wave pairings. Note that the $n=2$ CDI must be shifted in both the $x$- and $y$-directions; if we shift in only one direction then we have an odd number of true mirror symmetries, so the PH-symmetric system will have trivial sBZ topology.

The Bloch Hamiltonian for the shifted $n=2$ CDI is 
\begin{multline}
    \overline{h}_2=\sin2k_x\cos k_y\tau_x+\sin2k_y\cos k_x\tau_y\\+(m-\cos2k_x-\cos2k_y)\tau_z.
\end{multline}
The symmetries are \begin{align*}
    \overline{M}_x\overline{h}_2(k_x,k_y)\overline{M}_x^{-1}&=\overline{h}_x(-k_x+\pi,k_y)\\
    \overline{M}_y\overline{h}_2(k_x,k_y)\overline{M}_y^{-1}&=\overline{h}_x(k_x,-k_y+\pi),
\end{align*}
represented by $\overline{M}_x=\overline{M}_y=\tau_z$. There is also $C_2$ symmetry represented by $\tau_z$.

If a momentum-independent pairing described by $\hat\Delta=\Delta\tau_y$ is proximity-induced, the CDSC can realize a single intermediate phase characterized by $\mathcal{N}_r=\pm1$. A momentum-dependent, symmetry-breaking pairing such as $\hat\Delta=\Delta(\cos k_x+2\cos k_y)\tau_y$ will generate three intermediate phases, just like the $\sin k_x+2\sin k_y$ pairing for the unshifted case. By shifting the CDI in momentum space, we are able to realize emergent phases analogous to the unshifted case, but by inducing a simpler superconducting pairing.

\end{document}